\documentstyle[prc,epsf,aps]{revtex}
\begin{document}  
\title{Spin-dependent correlations and the semi-exclusive $^{16}$O$(e,e'p)$
reaction}
\author{Stijn Janssen, Jan Ryckebusch \footnote{corresponding author :
jan.ryckebusch@rug.ac.be}, Wim Van Nespen and Dimitri Debruyne}
\address{Department of Subatomic and Radiation Physics \protect\\
University of Gent, Proeftuinstraat 86, B-9000 Gent, Belgium}
\date{\today}
\maketitle

\begin{abstract}
The effect of central, tensor and spin-isospin nucleon-nucleon
correlations upon semi-exclusive $^{16}$O$(e,e'p)$ reactions is
studied for $Q^2$ and Bjorken $x$ values in the range $0.2 \lesssim
Q^2 \lesssim 1.1~$(GeV/c)$^2$ and 0.15~$\lesssim$~$x$~$\lesssim$~2.
The fully unfactorized calculations are performed in a framework that
accounts not only for the dynamical coupling of virtual photons to
correlated nucleon pairs but also for meson-exchange and
$\Delta_{33}$-isobar currents.  Tensor correlations are observed to
produce substantially larger amounts of semi-exclusive
$^{16}$O$(e,e'p)$ strength than central correlations do and are
predominantly manifest in the proton-neutron knockout channel.  With
the exception of the $x \approx 2$ case, in all kinematical situations
studied the meson-exchange and isobar currents are a strong source of
$A(e,e'p)$ strength at deep missing energies.  This feature gives the
$A(e,e'p)$ strength at deep missing energies a pronounced transverse
character.
\end{abstract}

\vspace{0.5cm}
\noindent
{\em PACS:} 25.30.-c,24.10.-i,25.30.Fj 

\noindent
{\em Keywords} : Nucleon-nucleon correlations, semi-exclusive $(e,e'p)$ 

\section{Introduction}
One of the major issues in ongoing electron scattering studies of
many-body nuclei is the study of short-range nucleon-nucleon ($NN$)
correlations.  At present, the major research efforts that aim at
studying $NN$ correlations with the aid of the electromagnetic probe
proceed along two lines.  The most direct source of information is the
simultaneous detection of the two (correlated) nucleons that are
ejected after the absorption of one single photon.  A more indirect
way of possibly probing the short-range correlations, is the
semi-exclusive $A(e,e'p)$ reaction provided that appropriate kinematic
regions are sampled and background contributions from two-body
currents \cite{kester96} and multiple-scattering effects can be kept
under control \cite{frankfurt97,ingoelba,demetriou99,amparo98}.
Numerous theoretical investigations that have addressed the
semi-exclusive $A(e,e'p)A-1$ process, have adopted a factorized
approach in which the electronuclear part and the information on the
energy and momentum distributions of nucleons in nuclei are strictly
separated in determining the differential cross sections
\cite{benhar99,morita99,claudio99}
\begin{equation}
\frac {d^6 \sigma} {d T_p d \Omega _p d \epsilon ' d \Omega _{\epsilon '}} =
\frac {p_p E_p} {(2\pi)^3}  \sigma _{ep} 
P_D \left( \vec{p}_m \equiv \vec{p}_p - \vec{q} , E_m \equiv \omega -
T_p - T_{A-1} \right) \; ,
\label{eq:factor} 
\end{equation}
here, the electronuclear part is contained in the factor $\sigma
_{ep}$ which is an off-shell extrapolation of the electron-proton
cross section.  The spectral function $P(\vec{k},E)$ contains the
nuclear-structure information and yields the joint probability
distribution of removing a nucleon with momentum $\vec{k}$ from the
target system and finding the residual $A-1$ nucleus at an excitation
energy $E$.  The distorted spectral function $P_D (\vec{k},E)$ that
appears in the above expression corrects $P(\vec{k},E)$ for
final-state interaction (fsi) effects which the ejectile undergoes.  A
factorized approach of the above form is subject to several
uncertainties.  Not only does it assume that the $NN$ ``correlations''
and final-state interactions are equally acting in the
longitudinal and transverse contributions to the cross sections, it
fails to include the strength from competing multi-body mechanisms
like for example meson-exchange currents.  There is accumulating
evidence for enhanced $A(e,e'p)$ strength of non-single particle
origin in the transverse response at deep missing energies
\cite{ulmer,lourie93,liyanage,maurik,dutta}.  The fact that a
comparable enhancement remains unobserved in the longitudinal channel
alludes to important physical phenomena that fall beyond the effects
implemented within factorized approaches based upon the
Eq.~(\ref{eq:factor}).

In Ref.~\cite{jan97} an unfactorized model for the calculation of
semi-exclusive $A(e,e'p)$ cross sections was outlined.  The model
included the effect of central (or, Jastrow) nucleon-nucleon
correlations as well as genuine two-body meson-exchange (MEC) and
isobar two-body currents (IC) that are often sources of $(e,e'p)$
strength that could be mistakingly interpreted as signals from the
short-range correlations.  Our method of calculating the
semi-exclusive $A(e,e'p)$ differential cross sections is based upon
the assumption that the short-range correlations will predominantly
manifest themselves as two-nucleon knockout even if only one of the
ejectiles is detected.  In this context we found it appropriate to set
up a calculational framework that aims at providing a unified
description of exclusive $A(e,e'NN)$ and semi-exclusive $A(e,e'p)$
processes.  In computing semi-exclusive $A(e,e'p)$ strength we
explicitly evaluate two-nucleon knockout cross sections and integrate
over the complete phase space of the undetected nucleons.  In calculating
the two-nucleon knockout cross sections we go beyond the mean-field
approach by implementing corrections for short-range nucleon-nucleon
correlations upon wavefunctions determined in a mean-field model.  In
this paper, our framework for computing semi-exclusive $A(e,e'p)$
strength is extended to include also spin-dependent nucleon-nucleon
correlations.  In addition, we address the issue how the effect of the
correlations manifests itself as a function of the Bjorken $x= \frac
{Q^2} {2 M_N \omega}$ scaling variable.  Moreover, we consider higher
momentum transfer regions which are now accessible at the Thomas
Jefferson National Accelerator Facility (TJNAF).  In doing this we aim
at finding out the most favorable conditions under which the MEC and
IC contributions to the semi-exclusive $A(e,e'p)$ processes can be
kept under control.  The outline of this paper is as follows.  The
basic assumptions underlying our $A(e,e'p)$ and $A(e,e'pN)$
calculations are sketched in Sections \ref{sec:eep} and
\ref{sec:eepN}.  Section \ref{sec:effectiveop} discusses the
spin-dependent correlations and the way they have been implemented in
the calculations.  Technical details are given in Section
\ref{sec:appendix}.  Results of the numerical $^{16}$O$(e,e'p)$
calculations for various kinematical regions in $Q^2$ ($0.2 \lesssim
Q^2 \lesssim 1.1~$(GeV/c)$^2$) and Bjorken $x$ values in the range
(0.15~$\leq$~$x$~$\leq$~2) are presented in Section \ref{sec:results}.
Finally, the summary and conclusions are given in Section
\ref{sec:conclusions}.

\section{Determination of two-nucleon knockout strength in correlated systems}
\label{sec:theory}

\subsection{The ``correlated'' part of the semi-exclusive $(e,e'p)$
cross section}
\label{sec:eep}
In determining the transition matrix elements that correspond to the
contribution from the $NN$ correlations to the
semi-exclusive $A(e,e'p)A-1$ reaction
\begin{equation}
\langle \Psi_f^ {A-1} \left( E_{A-1} \right) ; \vec{p}_p m_{s_p} 
\mid J_{\lambda}  \left( \vec{q} \right) \mid
\Psi_0^A \rangle \; ,
\end{equation}
we adopt the view that whenever a photon hits a ``correlated'' nucleon
both the latter and its ``correlated'' partner will be ejected
from the target system.  As a consequence, the $A-1$ system
for which the characteristics are determined in a semi-exclusive
$A(e,e'p)$ measurement is of the form
\begin{equation}
\mid \Psi_f^ {A-1} \left( E_{A-1} \right) ; \vec{p}_p m_{s_p}  \rangle
= \  \mid \Psi_f^{A-2}  \left( E_{A-2}, J_R M_R \right) ; \vec{p}_N m_{s_N} ;
\vec{p}_p m_{s_p} \rangle \; .
\end{equation}
Adopting such a reaction picture, the semi-exclusive $A(e,e'p)$ cross
section can be computed by integrating over the phase space of the
undetected but emitted nucleon
\begin{equation}
\frac {d^6 \sigma} {d T_p d \Omega _p d \epsilon ' d \Omega_
{  \epsilon' } } (e,e'p) = \sum _{N \equiv p,n} \int d \Omega_N d 
E_{A-2} \frac  
{d^9 \sigma} {d T_N d \Omega _N  d T_p d \Omega _p d \epsilon ' d
\Omega _{\epsilon '} } (e,e'pN) \; .
\end{equation}
Thus, the problem of computing semi-exclusive $A(e,e'p)$ cross
sections expands into the question of calculating two-nucleon knockout
cross sections.  This question is the subject of the next Subsection.

\subsection{Two-nucleon knockout cross section}
\label{sec:eepN}

In the standard fashion, the differential cross section for exclusive
$A(e,e'N_a N_b)A-2$ processes
\begin{equation}
A + e \left(\epsilon \right) \rightarrow \left(A-2 \right)
\left(E_{A-2}, \vec{p}_{A-2} \right) + N_a \left(E_a, \vec{p}_a
\right) + N_b  \left(E_b, \vec{p}_b \right) +  e \left(\epsilon'
\right) \; ,
\end{equation}
can be cast in the form
\begin{eqnarray}
\lefteqn{\frac{d^8 \sigma}{dT_b d\Omega_b d\Omega_a d\epsilon'
d\Omega_{ \epsilon'}} \left( e,e'N_a N_b\right) = \frac{1}{4 \left(  2
\pi \right)^8} p_a p_b E_a E_b f_{rec} \sigma_M } \nonumber \\
 & & \times \Biggl[ 
\left( \tan^2 \frac{\theta_e}{2} - \frac{1}{2}  \frac{ q_{ \mu}
q^{ \mu} }{ \vec{q}^{\ 2}} \right)
W_T \left( \theta_a , \theta_b, \phi _a - \phi _b
\right) + 
\left( \frac{ q_{\mu} q^{ \mu} }{ \vec{q}\ ^2 } \right)^2
W_L \left( \theta_a,  \theta_b, \phi_a - \phi_b \right)
\nonumber \\
 & & + 
\frac{q_{\mu} q^{\mu} }{ \sqrt{2} \mid \vec{q} \mid^3}
\left( \epsilon + \epsilon' \right) \tan \frac{\theta_e}{2}
W_{LT} \left( \theta_a , \theta_b, \phi _a - \phi _b , \frac {\phi _a
+ \phi _b} {2}  \right) +
\frac{1}{2} \frac{q_{\mu}q^{\mu}}{\vec{q}\ ^2}
W_{TT} \left( \theta_a, \theta_b, \phi _a - \phi_b, \frac {\phi _a
+ \phi _b} {2} \right) \Biggr] \; ,
\end{eqnarray}
where the Mott cross section $\sigma_M$ and the recoil factor
$f_{rec}$ are defined as
\begin{eqnarray} 
\sigma_M & = & \frac{e^4\cos^2\frac{\theta_e}{2}}{4 \epsilon^2 \sin^4
\frac{ \theta_e}{2}} \; ,\nonumber \\
f_{rec}^{-1} & = & 1 + \frac{E_a}{E_{A-2}} \left(1 - \frac{q \cos \theta_a }{
k_a} + \frac{k_b \cos \theta_{ab} }{k_a} \right)  \; .
\end{eqnarray}
Further, $(\theta _i,\phi _i)$ denote the polar and azimuthal angle of the
ejectile $i$  and $\theta_{ab}$ is the angle between the directions of the two
ejected nucleons.  The above differential cross section refers to the
``exclusive'' situation in which the residual $\left( A-2 \right)$ 
nucleus is created at a well-defined excitation energy.  We consider a
reference frame in which the momentum transfer $\vec{q}$ is aligned
along the $z$ axis and the $xz-$plane coincides with the electron
scattering plane.  The structure functions $W$ are expressed in terms
of the transition matrix elements for the two transverse and
longitudinal photon polarizations in the usual manner
\begin{eqnarray}
W_L \left( \theta_a , \theta_b, \phi _a - \phi_b \right) &=&
\sum_{m_{s_a}, m_{s_b},M_R} \left( m_{F}^{fi} \left( \lambda = 0
\right) \right)^{ \ast} \left( m_{F}^{fi} \left( \lambda = 0
\right) \right) \; , \\
W_T \left( \theta_a, \theta_b, \phi _ a - \phi_b \right) &=&
\sum_{m_{s_a}, m_{s_b},M_R} \biggl[ \left( m_{F}^{fi} \left
( \lambda = +1 \right) \right)^{\ast} \left( m_{F}^{fi} \left
(  \lambda = +1 \right) \right) \nonumber\\ 
& & + \left( m_{F}^{fi} \left( \lambda = -1 \right) \right)^{\ast}
\left( m_{F}^{fi} \left(  \lambda = -1 \right) \right) \biggr] \; ,\\
W_{LT} \left( \theta_a, \theta_b, \phi _a - \phi_b, \frac {\phi _a +
\phi_b} {2}  \right) &=& 2 \Re
\Biggl\{ \sum_{m_{s_a}, m_{s_b},M_R} \bigg[ \left( m_{F}^{fi}
\left( \lambda = 0 \right)  \right)^{  \ast} \left(  m_{F}^{fi}
\left(  \lambda = -1 \right) \right) \nonumber\\ 
& & - \left( m_{F}^{fi} \left( \lambda = 0 \right) \right)^{\ast}
\left( m_{F}^{fi} \left(  \lambda = +1 \right) \right) \biggr]
\Biggr\} \; , \\
W_{TT} \left( \theta_a, \theta_b, \phi _a - \phi_b, \frac{\phi _a +
\phi_b} {2}  \right)  & = & 2 \Re
\Biggl\{ \sum_{m_{s_a}, m_{s_b},M_R} \left( m_{F}^{fi} \left
( \lambda = -1 \right) \right)^{\ast} 
\left( m_{F}^{fi} \left( \lambda = +1 \right) \right)
\Biggr\} \; ,
\end{eqnarray}
where the sum extends over the spins of the three fragments in the
final state and the transition matrix elements are given by
\begin{eqnarray}
m_{F}^{fi} \left( \lambda = \pm 1 \right) & = & \langle \Psi_f^ {
A-2 } \left( E_{A-2}, J_R M_R \right) ; \vec{p}_a m_{s_a} ;
\vec{p}_b m_{s_b} \mid J_{\lambda= \pm1} \left( \vec{q} \right) \mid
\Psi_0^A \rangle, 
\nonumber \\ 
m_{F}^{fi} \left( \lambda = 0 \right) &=&
\langle \Psi_f^{A-2} \left( E_{A-2}, J_R M_R \right) ;
\vec{p}_a m_{s_a} ; \vec{p}_b m_{s_b} \mid \frac {q} {\omega}
J_{\lambda= 0} \left( \vec{q} \right) \mid \Psi_0^A \rangle \; .
\label{eq:transitionmat}
\end{eqnarray}
Here, $J_{\lambda = \pm1} \left( \vec{q} \right)$ ($J_{\lambda =0}
\left( \vec{q} \right)$) are the transverse
(longitudinal) components of the spatial nuclear current density. A convenient
and numerically stable way of dealing with the high dimensionality of  
transition matrix elements with three hadronic
objects ($A-2,N_a,N_b$) in the final state, is performing a partial
wave expansion for the ejectile's wave functions \cite{jana568}. This
method asks for 
a parallel multipole decomposition of the spatial current density
($J_{\lambda = 0, \pm1}$) in terms of the conventional Coulomb,
electric and magnetic operators
\begin{eqnarray}
J_{\lambda = 0} \left(\vec{q} \right) & = & 4 \pi \frac {\omega} {q}
\sum_{J=0}^{\infty} \widehat{J} \ i^J M_{J0}^{coul} \left(q\right) \;,
\nonumber \\
J_{\lambda = \pm 1} \left(\vec{q} \right) & = & - \sqrt{2 \pi} \sum_{J=1}^{\infty}
\widehat{J} \ i^J  \left[  T_{J \lambda}^{el} \left(q \right) + \lambda T_{J
\lambda}^{mag} \left(q \right) \right] \; ,
\end{eqnarray}
where $\widehat{J} \equiv \sqrt{ 2J + 1}$.
Each of the multipole components can be expressed as spherical tensor
operators according to
\begin{eqnarray}
M_{JM}^{coul} \left(q \right) &=& \frac {q} {\omega}  \int d \vec{r} \
j_J \left(qr \right) Y_{JM} \left( \Omega_r \right) 
J_{\lambda = 0} \left( \vec{r} \right) \; , 
\nonumber \\ 
T_{JM}^{el} \left(q \right) &=&
\frac{1}{q} \int d \vec{r} \ \vec{\nabla} \times \left[j_J \left(qr
\right) \vec{Y}_{J \left(J,1 \right)}^M \left( \Omega_r \right)
\right] \cdot \vec{J} \left ( \vec{r} \right) \; , 
\nonumber \\
T_{JM}^{mag} \left(q \right) &=& \int d \vec{r} \ j_J \left(qr \right)
\vec{Y}_{J \left(J,1 \right)}^M \left( \Omega_r \right) \cdot \vec{J}
\left( \vec{r} \right) \; .
\label{eq:multiop}
\end{eqnarray}

The transition matrixelements $m_{F}^{fi}$ contained in
Eq.~(\ref{eq:transitionmat}) can be cast in a closed form after
specifying the overlap between the final and initial state wave
functions. As outlined in Refs.~\cite{jana568,jan97}, within the
adopted framework that involves partial-wave expansions for the
nuclear wave functions and multipole decompositions for the operators,
the calculation of the cross sections with genuine two-body currents
(like MEC and IC) is eventually reduced to evaluating standard
two-body reduced matrix elements $ \langle a \;b ; J_1 \parallel
\widehat{O}_J \left( q \right) \parallel c \; d; J_2 \rangle $.  Here,
$\widehat{O}_J$ is one of the multipole operators from
Eq.~(\ref{eq:multiop}) and the quantum numbers $(a,b,c,d)$ refer to
either a bound $\left( a \equiv (n_a,l_a, \frac {1} {2}, j_a ) \right)
$ or a continuum eigenstate $\left( a \equiv (E_a, l_a, \frac {1} {2},
j_a) \right)$ of the mean-field potential.  In the forthcoming
Subsection we outline a technique that allows to evaluate the matrix
elements corresponding with the electromagnetic coupling to a
correlated nucleon pair.

\subsection{Ground- and final-state correlations in two-nucleon
knockout processes}
\label{sec:effectiveop}
Over the years, various techniques to correct independent particle
model (IPM) wave functions for
correlations have been developed.  All of these techniques, though,
face the complications that arise when going beyond a Slater
determinant approach.  In a correlated basis function (CBF) theory, the correlated
wave functions $\overline{ \Psi}$ are constructed by applying a
many-body correlation operator to the
uncorrelated wave functions $\Psi$ from a mean-field potential 
\begin{equation}
\mid \overline{ \Psi}\  \rangle =  \frac{1}{ \sqrt{N}}\ \widehat
{ {\cal G}} \mid  \Psi \ \rangle \; ,
\label{eq:coroperator}
\end{equation}
with the normalisation factor $ N \equiv \langle \ \Psi \mid \widehat{\cal
G}^{\dagger} \widehat{\cal G} \mid \Psi \ \rangle $.  The correlation
operator $ \widehat{\cal G}$ reflects a similar operatorial structure 
as the standard one-boson exchange parametrizations of the
nucleon-nucleon force and contains several terms
\begin{equation}
\widehat{\cal G} = \widehat{\cal S} \left[ \prod_{i<j=1}^A
\sum_{p} f^p (\vec{r}_{ij}) \widehat{O}_{ij}^p
\right] \; ,   
\end{equation}
where $\vec{r}_{ij} = \vec{r}_{i} - \vec{r}_{j}$ and $\widehat{\cal S}$
is the symmetrisation operator.  Apart from (relatively small) spin-orbit terms
the following operators are usually considered in constructing $\widehat {\cal G}$
\begin{equation}
\begin{array}{rclcrcl}
\widehat{O}_{ij}^{p=1} & = & 1 & 
& \widehat{O}_{ij}^{p=4}& = & \left( \vec{ \sigma}_i \cdot \vec
{ \sigma}_j \right) \left( \vec{ \tau}_i \cdot \vec{ \tau}_j \right)   \\
\widehat{O}_{ij}^{p=2}&=& \vec{ \sigma}_i \cdot \vec{ \sigma}_j &
& \widehat{O}_{ij}^{p=5}&=&\widehat{S}_{ij}  \\
\widehat{O}_{ij}^{p=3}&=& \vec{ \tau}_i \cdot \vec{ \tau}_j &\
& \widehat{O}_{ij}^{p=6}&=&\widehat{S}_{ij} \left( \vec{ \tau}_i \cdot
\vec{ \tau}_j \right) ,
\end{array}
\end{equation} 
where
$\widehat{S}_{ij}$ is the tensor operator $
\widehat{S}_{ij} = \frac{3}{r_{ij}^2} \left( \vec{\sigma}_i \cdot
\vec{r}_{ij} \right) \left( \vec{\sigma}_j \cdot \vec{r}_{ij} \right)
- \vec{\sigma}_i \cdot \vec{\sigma}_j$ .
Often, it has been reported that of all of the above components the
central (or Jastrow) $\left( p = 1 \right)$, the spin-isospin 
$\left( p = 4 \right)$ and the tensor $ \left( p = 6 \right)$ term cause the 
biggest correlation effects in the nuclear system
\cite{claudio99,pieper,benhar93,guardiola96}.  So, for the remainder of
the paper we consider a correlation operator of the form
\begin{equation}
\widehat{\cal G} = \widehat{ \cal S} \left[ \prod_{i<j=1}^A
\left( f^{p=1} \left( r_{ij} \right) + f^{p=4} \left( r_{ij}
\right)  \left( \vec{ \sigma}_i \cdot \vec{ \sigma}_j \right) \left
( \vec{ \tau}_i \cdot \vec{ \tau}_j \right) + f^{p=6} \left
( r_{ij} \right) \widehat{S}_{ij} \left( \vec{ \tau}_i \cdot \vec
{ \tau}_j \right) \right) \right] \; . 
\end{equation}
In many finite-nuclei calculations, only the (state-independent)
central correlations $\left( p = 1 \right)$ are retained.  In this
paper, we aim at developing a practical method in which both the
effect of the central and of the two major spin-dependent correlations
on two- and single-nucleon knockout processes can be evaluated.  It is
worth stressing that most of the cluster expansion techniques to
calculate the electromagnetic response of nuclei that are outlined in
literature refer to the ``inclusive'' case
\cite{pandharipande79,fantoni87}.  Hereby, closure properties are
maximally exploited and the computed responses usually refer to
responses integrated over all excitation energies of the final
system. For the present purposes, in which we aim at explicitly
determining the effect of the correlations on specific {\em exclusive}
and {\em semi-exclusive} channels, closure relations cannot be
exploited and the response to well-determined excited states of the
residual system has to be evaluated.  As a consequence, for our
purposes most of the cluster expansion techniques which are available
in literature are not directly applicable.  At the same time, the
scope of our theoretical calculations is different.  Indeed, our
calculational framework is meant to bridge the gap between the
many-body calculations that come up with predictions for the dynamical
correlation functions contained in $\widehat{\cal G}$ and exclusive
electronuclear reactions, where the predictions can be put to a test.

Usually, evaluating transition matrixelements between
correlated states is far from a trivial task.  We rely on an
``effective operator approach'' that was outlined in more detail in
Ref.~\cite{jan97}.  A formal formulation of the calculational scheme
was given by Feldmeier {\em et al.} in Ref.~\cite{feldmeier98} where
the method was introduced as the {\em Unitary Correlation Operator
Method (UCOM)}. In essence, the technique is based on an operator
expansion that is pursued after combining the correlation operators
acting on the nuclear states and the different terms contained in the
photon-nucleus interaction Hamiltonian.  Formally, this procedure
amounts to formally rewriting the transition matrix element between
{\sl correlated} states
\begin{equation}
\langle \ \overline{ \Psi}_f \mid \widehat{\Omega} \mid \overline
{ \Psi}_0 \rangle 
\end{equation}
as a transition between {\sl uncorrelated} states
\begin{equation}
\frac{1}{ \sqrt{N_0 N_f}} \langle \ \Psi_f
\mid \widehat{\Omega}^{eff}  \mid \Psi_0 \rangle  
\end{equation}
where the effect of the central, spin-isospin and tensor correlations
is implemented in an effective transition operator that combines the
effect of $NN$ correlations and the photon-nucleus Hamiltonian
\begin{eqnarray}
\widehat{\Omega}^{eff} &=& \widehat{\cal G}^{\dagger} \
\widehat{\Omega} \  \widehat{\cal G} \nonumber \\
 &=& \left( \prod_{i<j=1}^A \left[ 1 - \widehat{g} \left( i,j \right) + \widehat{s}
\left( i,j \right) + \widehat{t} \left( i,j \right) \right]  \right)^
{ \dagger}  \widehat{\Omega}  \left( \prod_{k<l=1}^A \left[ 1 - \widehat{g} \left( k,l \right) +
\widehat{s} \left( k,l \right)  + \widehat{t} 
\left( k,l \right)  \right] \right)\; .
\label{eq:lameff}
\end{eqnarray}
In the above expression the following shorthand notation for the
central, spin-isospin and tensor correlation operator is introduced
\begin{eqnarray}
\widehat{g} \left( i,j \right) & \equiv & 1 - f^{p=1} \left( r_{ij} \right) 
\\
\widehat{s} \left( i,j  \right) & \equiv & f^{p=4} \left( r_{ij}
\right) \left( \vec{ \sigma}_i \cdot \vec{ \sigma}_j \right) \left
( \vec{ \tau}_i \cdot \vec{ \tau}_j \right) 
\\
\widehat{t} \left( i, j \right) & \equiv & f^{p=6} \left( r_{ij}
\right) \widehat{S}_{ij} \left( \vec{ \tau}_i \cdot \vec{ \tau}_j
\right) \; .
\end{eqnarray}
Within the present context, the operator $\widehat{\Omega}$  stands for
the photon-nucleus interaction Hamiltoninan and contains 
apart from the standard one-body part of the impulse approximation
also two-body terms 
\begin{equation}
\widehat{ \Omega} \equiv \sum_{i=1} ^{A} \widehat{ \Omega}^{ \left[ 1 \right]}
\left( i \right) + \sum_{i<j=1} ^{A} \widehat{ \Omega}^{ \left[ 2 \right]}
\left( i,j \right) \; . 
\end{equation}
After expanding the effective transition operator of
Eq.~(\ref{eq:lameff}) and retaining solely the
terms that are linear in any of the correlation operators
$\widehat{g}, \widehat{s}$ or $\widehat{t}$ one obtains an expression of
the type
\begin{equation}
\widehat{ \Omega}^{eff} \approx    \left( \sum_{i=1} ^{A}  \widehat{ \Omega}^{ \left[ 1
\right]} \left( i \right) + \sum_{i<j=1} ^{A} \widehat{ \Omega}^{ \left[ 2
\right]} \left( i,j \right) \right) + 
\left( \widehat{ \Omega}^{ \left[ 1 \right] , in} +\  \widehat
{ \Omega}^{ \left[ 2 \right] , in}  +\ \widehat{ \Omega}^{ \left
[ 1 \right] ,fi} +\  \widehat{ \Omega}^{ \left[ 2 \right] ,fi} \right)
\; ,
\label{eq:effop}
\end{equation}
where the index $in$  $(fi)$ refers to the initial (final) state
correlations and the following operators were introduced 
\begin{eqnarray}
\widehat{ \Omega} ^{ \left[ 1 \right] ,in} &=& \sum_{i<j} \left
[ \widehat{ \Omega}^{ \left[ 1  \right]} \left( i \right) +  \widehat
{ \Omega}^{ \left[ 1 \right]}  \left( j \right) \right] \widehat{l}
\left( i,j \right) \nonumber \\ 
 & &  + \sum_{i<j<k} \left[ \widehat{ \Omega}^{ \left[ 1 \right]}
\left( i \right)  \widehat{l} \left( j,k \right) + \widehat
{ \Omega}^{ \left[ 1 \right]} \left( j \right)  \widehat{l} \left
( i,k \right) +  \widehat{ \Omega}^{ \left[ 1 \right]} \left( k
\right)  \widehat{l}  \left( i,j \right) \right] \;, \nonumber \\
 & &  \nonumber \\
\widehat{ \Omega} ^{ \left[ 2 \right] ,in} &=& \sum_{i<j} \;
\widehat{  \Omega}^{ \left[ 2 \right]} \left( i,j \right)  \widehat{l}
\left( i,j \right)  \nonumber \\
 & & + \sum_{i<j<k} \left[ \widehat{ \Omega}^{ \left[ 2 \right]}
\left( i,j \right) \widehat{l}  \left( i,k
\right) + \widehat{ \Omega}^{ \left[ 2 \right]} \left( i,j \right)
\widehat{l}  \left( j,k \right) +  \widehat{ \Omega}^{ \left[ 2
\right]} \left( i,k \right)  \widehat{l}  \left( i,j \right)
\right.  \nonumber \\
 & & + \left. \widehat{ \Omega}^{ \left[ 2 \right]} \left( i,k \right)
\widehat{l}  \left( j,k \right) +  \widehat{ \Omega}^{ \left[ 2
\right]} \left( j,k \right)  \widehat{l} \left( i,j \right) +
\widehat{  \Omega}^{ \left[ 2 \right]} \left( j,k \right)  \widehat{l}
\left( i,k \right) \right]   \nonumber \\
 & & + \sum_{i<j<k<m} \left[ \widehat{ \Omega}^{ \left[ 2 \right]}
\left( i,j \right) \widehat{l}  \left( k,m \right) +  \widehat
{ \Omega}^{ \left[ 2 \right]} \left( i,k \right)  \widehat{l}  
\left( j,m \right) +  \widehat{ \Omega}^{ \left[ 2 \right]} \left( i,m
\right)  \widehat{l}  \left( j,k \right)  \right. \nonumber \\
 & & + \left. \widehat{ \Omega}^{ \left[ 2 \right]} \left( j,k \right)
\widehat{l}  \left( i,m \right) +  \widehat{ \Omega}^{ \left[ 2
\right]} \left( j,m \right)  \widehat{l} \left( i,k \right) +
\widehat{ \Omega}^{ \left[ 2 \right]} \left( k,m \right)  \widehat{l}
\left( i,j \right) \right], \nonumber  \\
 & & \nonumber \\
\widehat{ \Omega} ^{ \left[ 1 \right] ,fi}  &=&  \left( \widehat{ \Omega} ^
{ \left[ 1 \right] ,in} \right)  ^{\dagger}  
\nonumber \\
\widehat{ \Omega} ^{ \left[ 2 \right] ,fi}  &=&  
\left( \widehat{ \Omega} ^{ \left[ 2 \right] ,in}  \right)
^{\dagger} \; .
\end{eqnarray}
The operator $\widehat{l} $ is a shorthand notation for the sum
of the central, spin-isospin and tensor correlation operator
\begin{equation}
\widehat{l} \left( i,j \right) = 
- \widehat{g} \left( i , j \right)  
+ \widehat{s} \left( i , j \right) 
+ \widehat{t} \left( i , j \right) \; .
\end{equation}
In the absence of initial and final-state correlations only the first
two terms in the expansion of Eq.~(\ref{eq:effop}) would not vanish.
At large internucleon distances ($r_{ij} \geq$ 4 fm) we have
$\widehat{l}(i,j) \longrightarrow 0$ and the operator $\widehat {
\Omega}^{eff}$ heals to the uncorrelated operator $\widehat{
\Omega}$.  It is worth remarking that the ground- and
final-state correlations are to be treated on the same footing in
order to guarantee that the effective operator formalism produces
hermitian transition operators.

Even in the lowest-order expansion of  
Eq.~(\ref{eq:effop}) the effective operator 
$ \widehat{ \Omega}^{eff} $ contains up to four-body operators.  In
exclusive two-nucleon knockout processes that sample two-body kinematics 
the three- and four-body operators are expected to produce very small
corrections.  Additionally, the probability of finding three and more
correlated nucleons at normal nuclear densities is generally conceived
to be extremely small.  In the ``single pair approximation'' (SPA),
all terms up to
two-body operators in the expansion of Eq.~(\ref{eq:effop}) are retained.
Substituting the one-body hadronic current $J_{\lambda}^{\left[ 1
\right]}$ for $\widehat{\Omega}^{\left[ 1 \right]}$ and the two-body
current $J_{\lambda}^{\left[ 2 \right]}$ for $\widehat{\Omega}^{\left[
2 \right]}$ one is left with the following effective transition
operator 
\begin{equation}
\widehat{ \Omega}_{\lambda}^{eff}  =  \left\{ \sum_{i=1}^A
J_{\lambda} 
^{ \left[ 1 \right] } \left( i \right) + \sum_{ i<j=1}^A J_{\lambda}^{ \left[ 2
\right]} \left( i,j \right)  \right\}
+ \sum_{ i<j=1}^A  J_{\lambda}^
{ \left[ 1 \right],in }  \left( i,j \right) + \sum_{ i<j=1}^A  J_{\lambda}^
{ \left[ 1 \right],fi }  \left( i,j \right) \; \; \; \; \;   
\left( \lambda \! = \! 0, \pm 1 \right) \;  .
\label{eq:spa}
\end{equation} 
The physical interpretation of the various terms appearing in
Eq.~(\ref{eq:spa}) is straightforward.  The first two terms reflect
the electromagnetic interaction hamiltonian for uncorrelated
mean-field states. The $J_{\lambda}^{ \left[
2 \right]} \left( i,j \right)$ are the mesonic and isobaric currents
for which our model assumptions can be found in
Refs~\cite{jan97,jan99}.  For the one-body charge and current density
operator we consider the standard form of the non-relativistic impulse
approximation.  Then, the ``correlation'' terms 
that occur in Eq.~(\ref{eq:spa}) are effective two-body operators and read
\begin{eqnarray}
J_{\lambda=0}^{ \left[ 1 \right],in} \left( i,j \right) 
&=&  e \frac{\omega} {q} \biggl[ \delta \! \left( \vec{r} - \vec{r}_i
\right)  \frac{1+ 
\tau_{z,i}}{2} +   \delta \! \left( \vec{r} - \vec{r}_j \right)
\frac{1+ \tau_{z,j}}{2} \biggr] 
\left[\widehat{s} \left( i,j
\right) + \widehat{t} \left( i,j \right) - \widehat{g} \left( i,j \right)
\right] \;,  \nonumber \\  
\vec{J}_{\lambda= \pm1}^{\  \left[ 1 \right],in} \left( i,j \right)
& = & \frac{ e \hbar}{ 2 M_N}  \Biggl[ \frac{1+ \tau_{z,i}}{2i} 
\left[ \vec{\nabla}_i , \delta  \left( \vec{r} - \vec{r}_i \right) \right]_+
+ \frac{1+ \tau_{z,j}}{2i} \left[ \vec{ \nabla}_j,  \delta 
\left( \vec{r} - \vec{r}_j \right) \right]_+  
\nonumber \\
& & +  \mu_i \delta  
\left( \vec{r} - \vec{r}_i \right) \vec{  \nabla}_i \times \vec{ \sigma}_i
+  \mu_j \delta \! \left( \vec{r} - \vec{r}_j \right) \vec
{  \nabla}_j \times \vec{ \sigma}_j   \Biggr] 
\left[
\widehat{s} \left( i,j \right) + \widehat{t} \left( i,j \right)
- \widehat{g} \left( i,j \right) \right] \;, \nonumber \\
\vec{J}_{\lambda}^{\  \left[ 1 \right],fi} & = & 
\left(\vec{J}_{\lambda}^{\  \left[ 1 \right],in} \right)^{\dagger} \; ,
\label{eq:jeff}
\end{eqnarray}
where we have introduced the charge density operator in the
longitudinal component of the vector current, to impose current
conservation at the one-body level.  The anticommutator
$\left[\widehat{A},\widehat{B} \right]_+ \equiv
\widehat{A}\widehat{B}+\widehat{B}\widehat{A}$ is defined in the
standard fashion. The reduced matrix elements corresponding with the
effective currents operators of Eq.~(\ref{eq:jeff}) are given in
Appendix A.  It is worth stressing that the tensor term can only be
evaluated at a large computational cost.  Strictly speaking, the
effective operator of Eq.~(\ref{eq:spa}) has an additional term of the
form $\sum_{i<j} \left( \widehat{\Omega}^{[2]} \left(i,j\right) \widehat{l}
\left(i,j\right) + h.c. \right)$.  This term regularizes the two-body currents
for dynamical correlations in the nuclear wave functions. The
short-range behaviour of two-body currents is commonly regularized in
a semi-phemonological manner through introducing hadronic form factors
of the monopole form.  In our calculations we have adopted this
procedure with a cut-off formfactor of $\Lambda _\pi$=1250~MeV.  The
cumulative effect of hadronic form factors, which seriously cut on the
short-range part of the two-body currents, and dynamical short-range
correlations in the wave functions was shown to be small
\cite{miller76,marcvdh} and is neglected in our calculations.
Further, we have adopted electromagnetic nucleon form factors in the
standard dipole parametrization.

\section{Results of Numerical Calculations}
\label{sec:results}

\begin{figure}
\begin{center}
{\mbox{\epsfysize=14.cm\epsffile{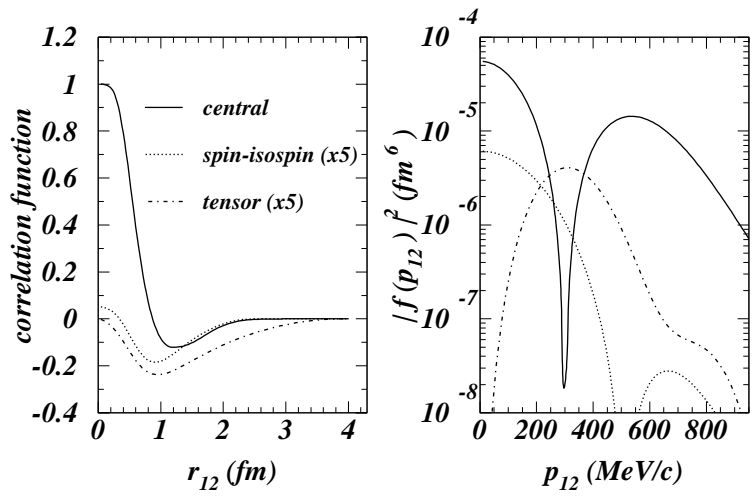}}}
\end{center}
\caption{The correlation functions used in the presented
calculations. For the coordinate space representation, the
spin-isospin and tensor correlation function are multiplied by a
factor of five.}
\label{fig:corfunc}
\end{figure}

In the forthcoming, $^{16}$O$(e,e'p)$ results will be presented.
Although our methods are applicable to all even-even target nuclei, we
have selected a light target nucleus for several reasons.  A first
reason is that the amount of computing time required to determine the
differential cross sections, depends on the number of nucleon pairs
that can be formed in the target nucleus.  Second, unlike for heavy
target nuclei, for a $Z=8$ nucleus Coulomb distortion effects in the
electron waves can be safely ignored.  A third consideration concerns
the fact that light nuclei are far more transparent for nucleon
emission than heavy nuclei \cite{abbott}.  As a consequence, the
attenuation of the emitted protons and neutrons is a serious
complication in heavy nuclei.  After all, the physics behind $NN$
``correlations'' is conceived to be rather mass independent and more
favorable conditions to study them with the electromagnetic probe are
probably reached when selecting light target nuclei.  In constructing
the Slater determinants for the initial and final states we use a
realistic set of single-particle wave functions obtained from a
Hartree-Fock calculation with an effective Skyrme force.  The use of
other sets of realistic single-particle wave functions does not
significantly alter our conclusions.  As both the ground-state wave
function for the target nucleus $^{16}$O and the scattering states for
the two ejectiles are constructed from the same Hartree-Fock
potential, the transition matrix elements are free from spurious
contributions related to non-orthogonality deficiencies.  In
implementing the effects of ground-state correlations, we use the
central correlation function $g$ from the G-matrix calculations in
nuclear matter with the Reid potential by W.H. Dickhoff and
C. Gearhart \cite{gearhart}.  This correlation function is shown in
Figure~\ref{fig:corfunc} and heals to zero at $r_{12} \geq$~2.5~$fm$.
It has a hard core at short internucleon distances $r_{12}$,
guaranteeing that the nucleons repel each other strongly enough when
they come too close.  Using this correlation function, we achieved a
fair agreement with the available data sets for $^{12}$C$(e,e'pp)$
\cite{blom98} and $^{16}$O$(e,e'pp)$ \cite{jan99,gercoprl2,ronald99}.
For the spin-isospin and tensor correlation functions we use those
that were obtained by S.~Pieper {\em et al.} in a variational
calculation for the ground state of $^{16}$O with the Argonne $v_{14}$
NN potential \cite{pieper}.  It is worth stressing that the
correlation functions are conceived to constitute a general feature of
atomic nuclei and that the correlation functions are predicted to
exhibit a very small $A$ dependence.  From Fig.~\ref{fig:corfunc} it
is clear that in coordinate space the spin-isospin and tensor
correlation function appear considerably weaker than the central one.
In momentum space, though, a very different picture emerges and for
momenta in the range 200-400~MeV/c the correlation functions related
to the spin-dependent terms are of comparable magnitude than the
central correlation function.  On the basis of the behaviour of the
correlation functions in momentum space, one could expect important
contributions from the tensor correlations at intermediate missing
(or, relative pair) momenta.

\begin{figure}
\begin{center}
{\mbox{\epsfysize=19.cm\epsffile{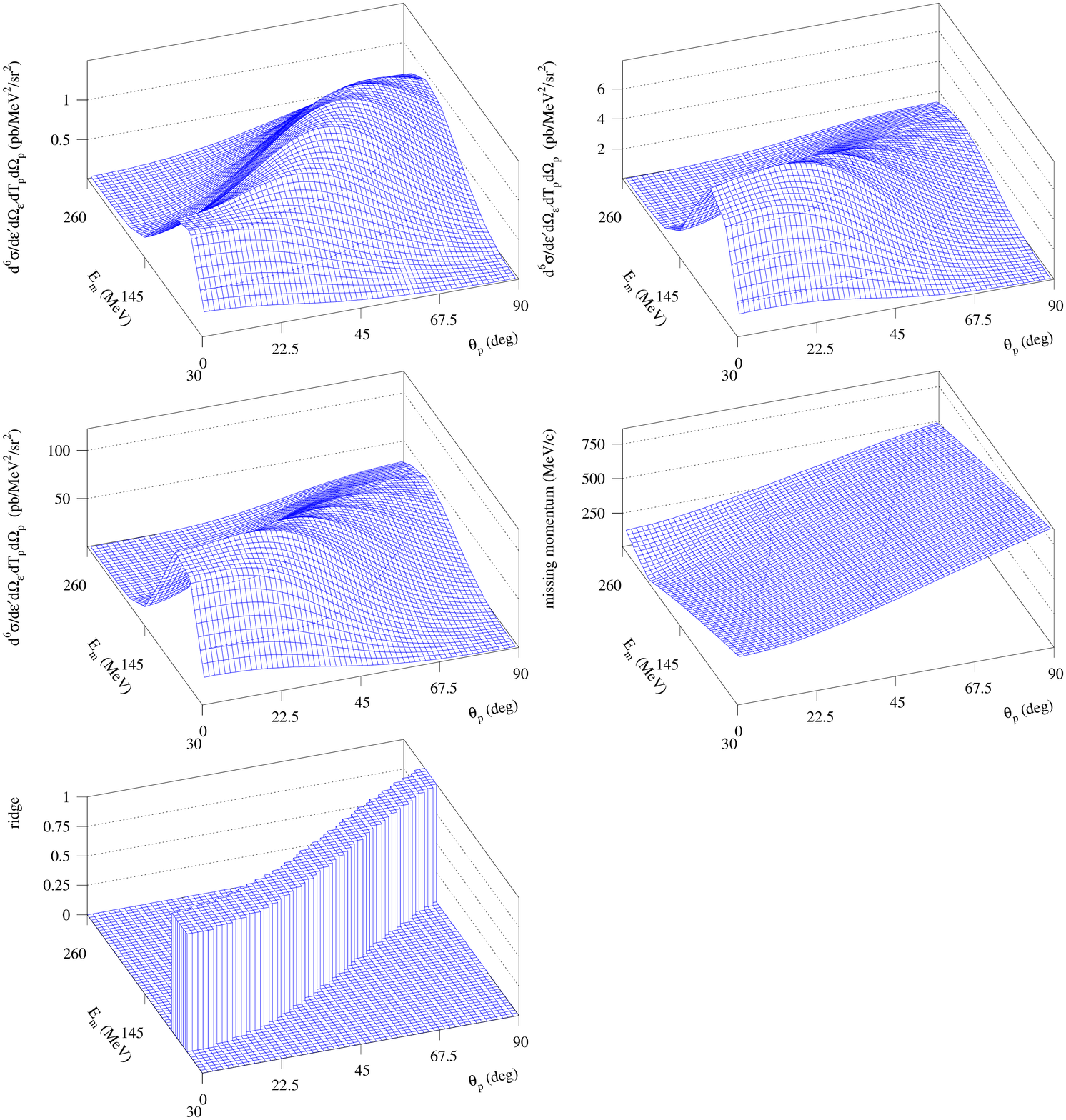}}}
\end{center}
\caption{The calculated contribution from two-nucleon knockout to the
differential $^{16}$O$(e,e'p)$ cross section versus missing energy and
proton angle at is $\epsilon$=1.2~GeV, $\epsilon '$=0.9 ~GeV and
$\theta_e$=16$^o$ (or, $x \approx 0.15 $ and $q = 0.42$~GeV/c). The
upper left panel includes solely the central correlations and the
upper right panel has both central and tensor correlations.  The lower
left panel, on the other hand, includes apart from the central and
tensor correlations also the MEC and IC.  The variation of the missing
momentum versus missing energy and proton angle is shown in the lower
right panel. The lower left panel shows the position of the ``ridge''
in the $(E_m,p_m)$ plane.  Hereby, the variable $<E_x^{hh'}>$ was
varied between 0. and 40.~MeV.}
\label{fig:central}
\end{figure}

\begin{figure}
\begin{center}
{\mbox{\epsfysize=13.cm\epsffile{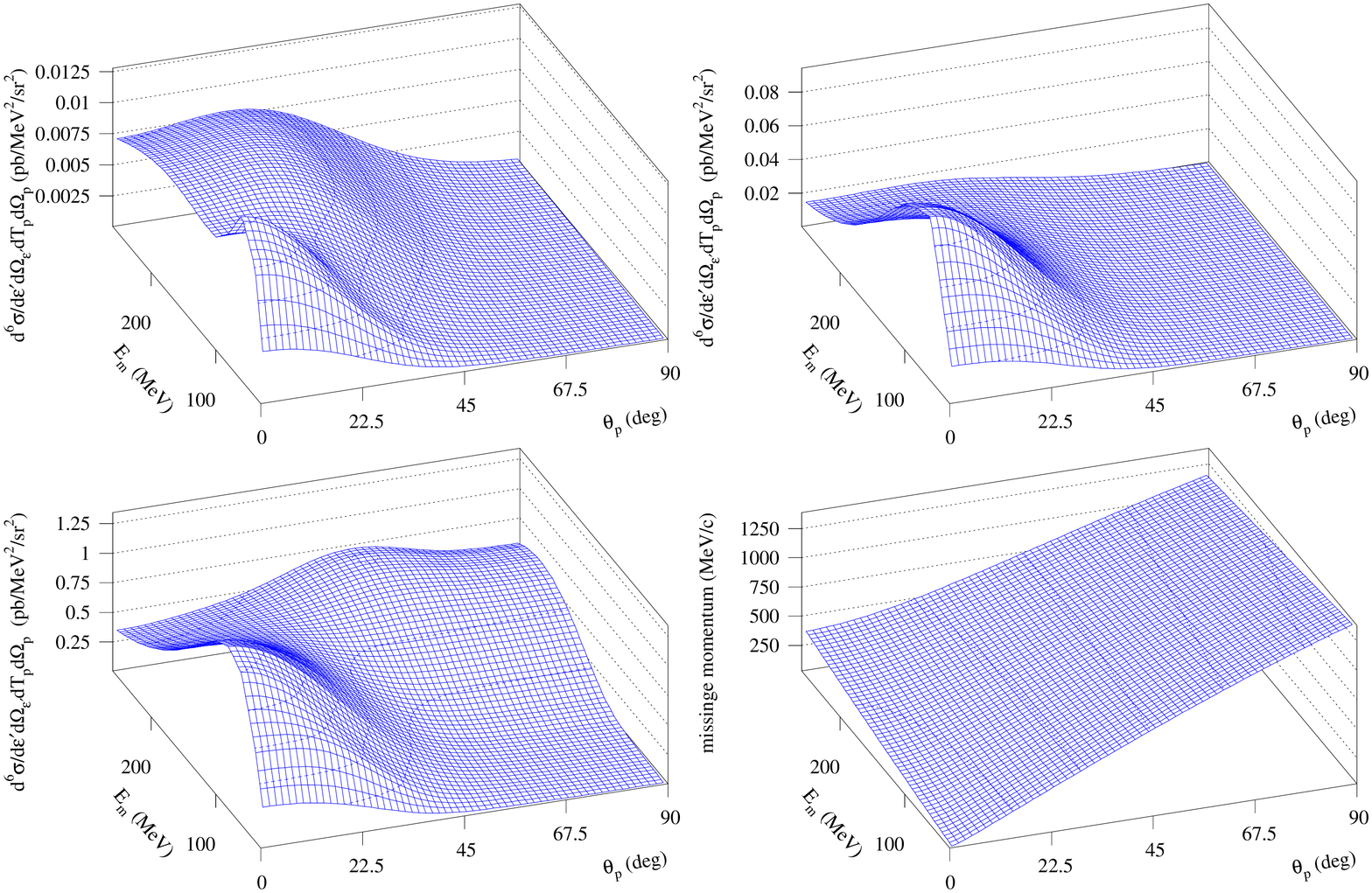}}}
\end{center}
\caption{The calculated contribution from two-nucleon knockout to the
differential $^{16}$O$(e,e'p)$ cross section versus missing energy and
proton angle at $\epsilon$=2.442~GeV, $\epsilon '$=1.997 ~GeV and
$\theta_e$=23.4$^o$ (or, $x \approx 1$ and $q = 1.0$~GeV/c). The upper
left panel includes solely the central correlations and the upper
right panel has both central and tensor correlations.  The lower left
panel, on the other hand, includes apart from the central and tensor
correlations also the MEC and IC.  The variation of the missing
momentum versus missing energy and proton angle is shown in the lower
right panel.}
\label{fig:centralx1}
\end{figure}

\begin{figure}
\begin{center}
{\mbox{\epsfysize=13.cm\epsffile{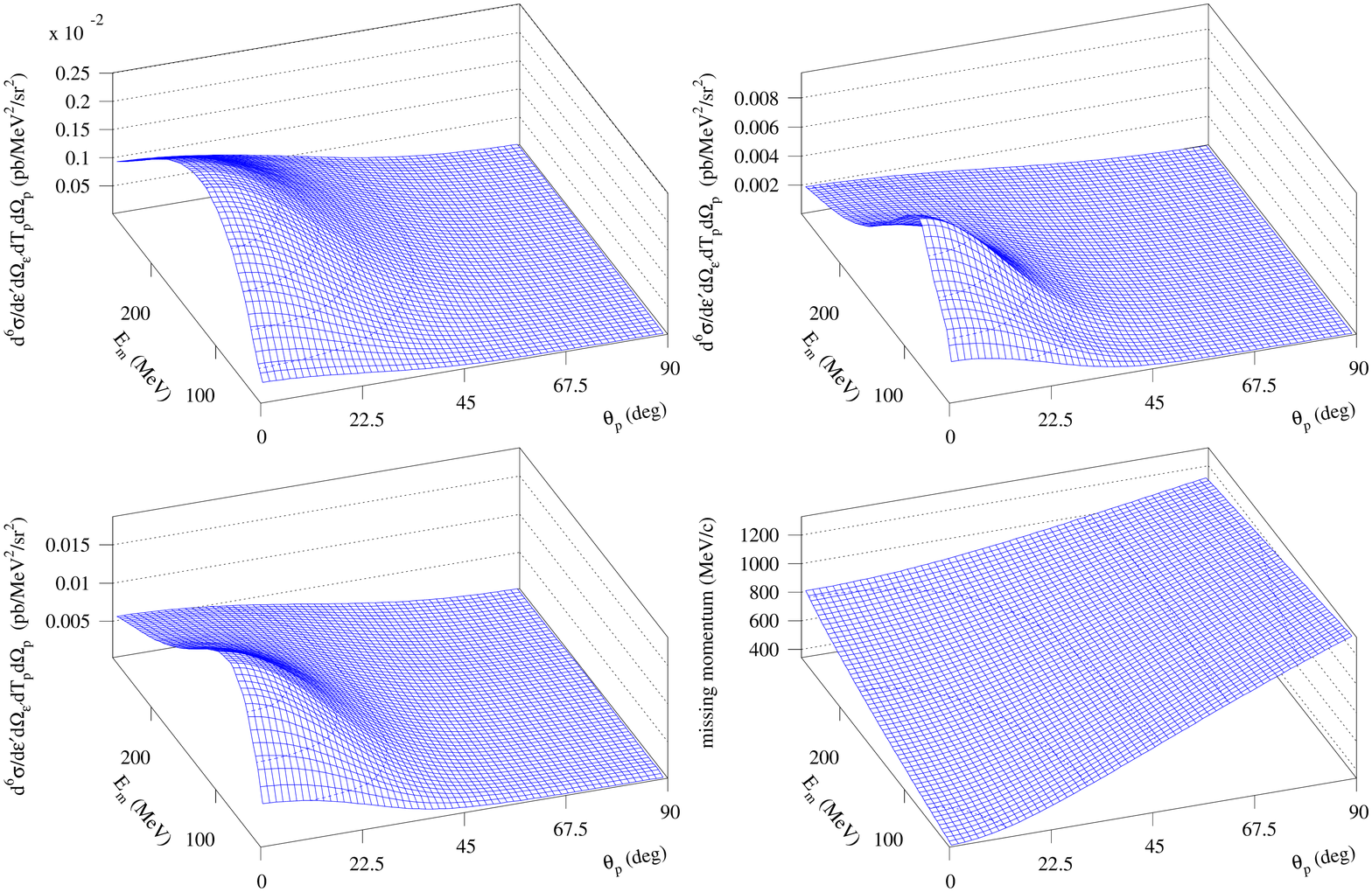}}}
\end{center}
\caption{The calculated contribution from two-nucleon knockout to the
differential $^{16}$O$(e,e'p)$ cross section versus missing energy and
proton angle at $\epsilon$=2.5~GeV, $\epsilon '$=2.2 ~GeV and
$\theta_e$=26.0$^o$ (or, $x \approx 2$ and $q = 1.1 GeV/c$).  The
upper left panel includes solely the central correlations and the
upper right panel has both central and tensor correlations.  The lower
left panel, on the other hand, includes apart from the central and
tensor correlations also the MEC and IC.  The variation of the missing
momentum versus missing energy and proton angle is shown in the lower
right panel.}
\label{fig:centralx2}
\end{figure}

All the $^{16}$O$(e,e'p)$ differential cross sections presented below
are obtained by incoherently adding the separately computed
$^{16}$O$(e,e'pn)$ and $^{16}$O$(e,e'pp)$ strengths.  In this
procedure, two-nucleon knockout from all possible shell-combinations
(viz. $(1s1/2)^2, (1s1/2)(1p3/2), (1s1/2)(1p1/2), (1p1/2)^2,
(1p1/2)(1p3/2)$ and $(1p3/2)^2$) is incorporated.  We have carried out
$^{16}$O$(e,e'p)$ calculations for three kinematical conditions.  The
first corresponds with a small Bjorken scaling variable $x \approx
0.15 $, the second with $x \approx 1$ (quasi-elastic conditions) and
the third with $x \approx 2$ where some typical quasi-deuteron like
structures are expected to occur.  The quasi-elastic case coincides
with the kinematics of a recent TJNAF experiment (E89-003)
\cite{liyanage}.  Figures \ref{fig:central}-\ref{fig:centralx2}
summarize the results of the $^{16}$O($e,e'p$) calculations.  We first
address the issue how the central, tensor and spin-isospin
correlations manifest themselves in the $(e,e'p)$ differential cross
sections. For each of the three kinematical regions under
consideration we study the differential $^{16}$O$(e,e'p)$ cross
section versus missing energy and proton angle $\theta_p$ in planar
kinematics ($\phi _p = 0^o$).  The observed functional dependence in
$\theta _p$ allows to study the missing momentum variations of the
cross sections.  For the sake of convenience, for each electron
kinematics we have added a panel with the variation of the missing
momentum versus missing energy and proton angle.  Roughly, the probed
missing momentum $p_m$ increases with increasing $\theta_p$ and
missing energy $E_m$.  It is clear that by measuring the missing
energy dependence of the $(e,e'p)$ cross section for a number of
proton polar angles, access to broad regions in the $(E_m,p_m)$ plane
is gained. Comparing the two upper panels in Figures
\ref{fig:central}, \ref{fig:centralx1} and \ref{fig:centralx2}, a
common qualitative feature emerges.  For all three considered regions
in the Bjorken scaling variable $x$ the strength attributed to the
tensor correlations largely overshoots the strength from the central
correlations.  This holds in particular for the small proton angles
$\theta _p$.  At these angles, typically the smallest missing momenta
are probed.  The effect of the spin-isospin correlations is found to
be at the few percent level, and, therefore, negligible.  The central
correlations are observed to manifest themselves in a wider range of
the ($E_m,\theta_p$) space than the tensor correlations do.  The
effect of the tensor correlations is seemingly confined to a region of
relatively low and moderate missing momenta, whereas the contribution
from the central correlations extends to higher proton angles $\theta
_p$ where typically higher missing momenta are probed.  This
qualitative behaviour of the calculated differential cross sections reflects the
fact that central correlations are the most important correlations in
the spectral function at really high missing momenta, whereas the
intermediate range is usually dominanted by the tensor correlations
\cite{dimitrienlieven}.  Such a qualitative behaviour can also be
inferred from Fig.~\ref{fig:corfunc}.  Another interesting feature of
how the ground-state correlations manifest themselves in the
$(E_m,p_m)$ plane is that the peak of the $^{16}$O$(e,e'p)$
differential cross sections shifts to higher missing energies $E_m$ as
one gradually moves out of parallel kinematics and higher $\theta_p$
angles (and consequently, missing momenta) are probed.  This
observation is a manifestation of a well-known feature of the
correlated part of the spectral function.  Indeed, the average missing
energy $\left< E_m \right>$ is predicted to increase quadratically in
the missing momentum : $ \left< E_m \right> = \frac {A-2} {A-1} \frac
{p_m^2}{2M_N} + S_{2N} + <E^{hh'}_{A-2}>$, where $S_{2N}$ is the threshold
energy for two-nucleon knockout and $<E_{A-2}^{hh'}>$ the average
excitation energy of the $A-2$ system that was created after two
nucleons escaped from the orbits characterized by the quantum numbers
$h$ and $h'$.  Moreover, the strength in the correlated part of the
spectral function is often predicted to be localized in a rather
narrow region on both sides of this central $E_m$ value (the so-called
``ridge'' in the spectral function).  The upper panels in
Figs.~\ref{fig:central}-\ref{fig:centralx2} clearly illustrate that
the largest fraction of the calculated $^{16}$O$(e,e'p)$ strength is
localized on both sides of the ``ridge''.  For the sake of
convenience, a panel with the exact location of the ridge in the
$(E_m,p_m)$ plane was added to Figure~\ref{fig:central}.  The
calculated strength appears in a rather wide band of missing energies
around this ridge, though.  Despite the fact that our calculations are
unfactorized in nature, the above observations with regard to the
qualitative behaviour of the calculated differential cross sections,
illustrate that our unfactorized results exhibit the same qualitative
features than what could be expected to happen in a factorized
approach based upon the Equation~(\ref{eq:factor}) using a realistic
``correlated'' spectral function.

One of the major advantages of our unfactorized approach to the
semi-exclusive $A(e,e'p)$ reaction, is that also the strength from the
MEC and IC can be evaluated in the same calculational framework in
which also the contributions from ground-state correlations are
determined.  This allows to calculate some of the additional
reaction-mechanism effects that break the direct link between the
measured cross sections $(e,e'p)$ differential cross sections and the
spectral function. The predictions for the differential cross sections
when including both the ground-state correlations and the MEC/IC
currents are contained in Figs.~\ref{fig:central}-\ref{fig:centralx2}.
For the low $x$ kinematics, the calculated contribution from the MEC
and IC overshoots the calculated strength from the $NN$
correlations. This confirms earlier findings for the $^{12}$C$(e,e'p)$
\cite{kester96} and the $^{4}$He$(e,e'p)$ \cite{leeuwe98} reaction.
Even in the quasi-elastic situation ($x \approx 1$) of
Fig.~\ref{fig:centralx1} the calculated $^{16}$O$(e,e'p)$ strength
that is created through MEC and IC is substantially larger than the
strength attributed to the nucleon-nucleon correlations.  This is
particularly the case at backward nucleon angles $\theta _p$ where the
two-body currents dominate the $(e,e'p)$ strength in the $(E_m, \theta
_p)$ plane. An interesting feature is observed for the low $x$ case of
Fig.~\ref{fig:central}.  The MEC/IC create almost an order of
magnitude more proton-knockout strength than the $NN$ correlations do.
A striking observation, though, is that in this particular case the
MEC/IC generate the $(e,e'p)$ strength in the same regions of the 
$(E_m,\theta_p$) plane as the $NN$ correlations do.  Thus, the mere
observation of a ``ridge'' in the continuous part of the $(e,e'p)$
spectrum does not necessarily imply that signals directly pointing to
the $NN$ correlations are detected.  Better conditions to isolate the
$NN$ correlations with the aid of the $(e,e'p)$ reaction, are
predicted for the $x \approx 2$ case (Fig.~\ref{fig:centralx2}).
Despite the fact that in the considered kinematics reasonably high
missing momenta ($p_m \geq 400$~MeV/c) are probed, the role played by
the MEC and IC is moderate in comparison with the role played by the
nucleon-nucleon correlations.  It should be stressed that the $x
\approx 2$ region is only accessible at facilities with sufficient
inital electron energy, like TJNAF, and implies relatively small cross
sections (for the typical example considered for the results of
Fig.~\ref{fig:centralx2} the cross sections are of the order
0.01~pb/MeV$^2$/sr$^2$).  A recent account of the $(e,e'p)$ program at
TJNAF is given in Ref.~\cite{templon99}.  An accepted proposal to
measure $^{12}$C$(e,e'p(N))$ at $x \approx 2$ at TJNAF kinematics is
described in Ref.~\cite{shalev}.

\begin{figure}
\begin{center}
{\mbox{\epsfysize=9.cm\epsffile{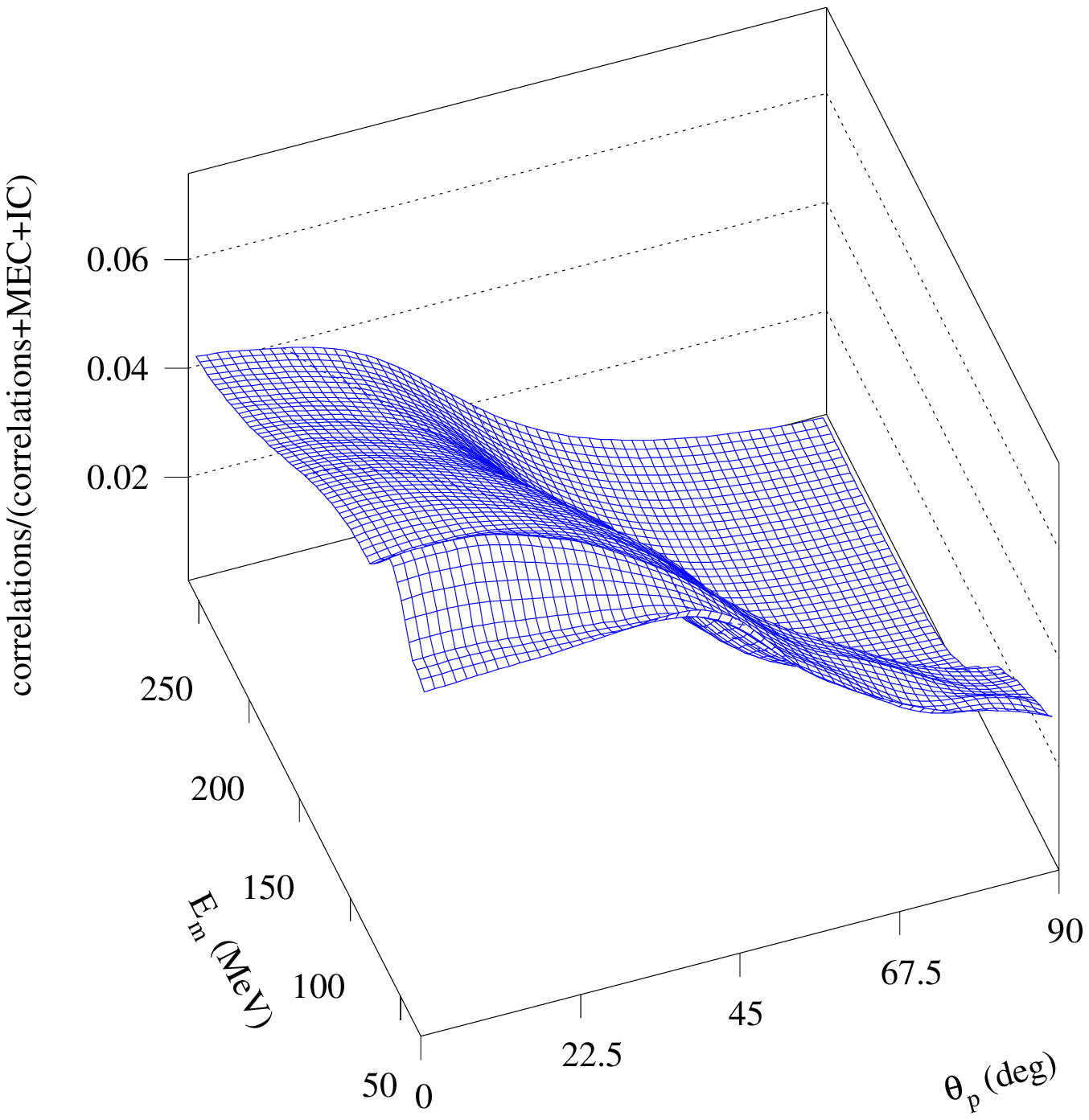}}}
{\mbox{\epsfysize=9.cm\epsffile{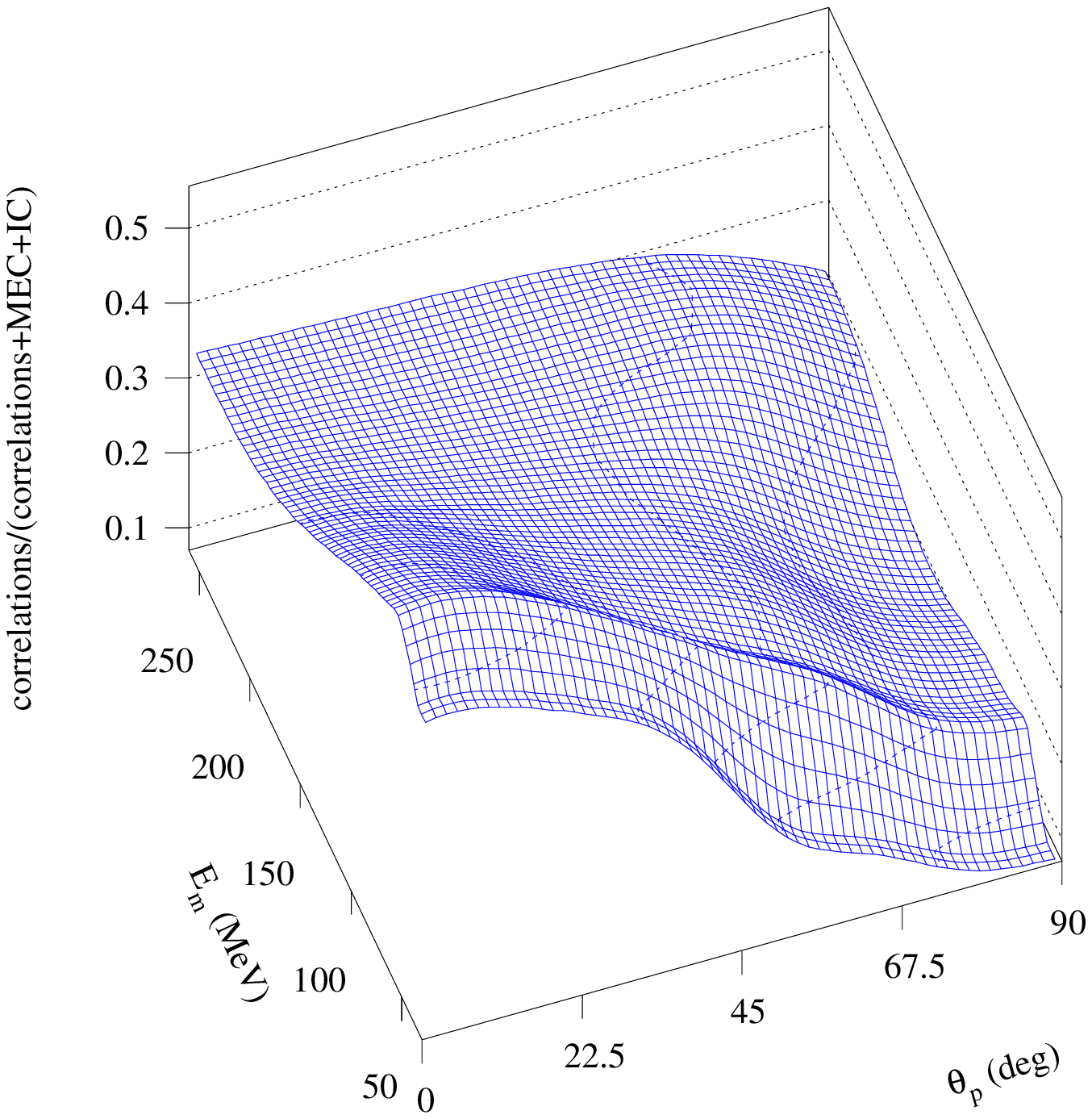}}}
\end{center}
\caption{The ratio of the strength from the nucleon-nucleon
correlations to the full semi-exclusive $^{16}$O$(e,e'p)$ strength
(including MEC and IC) versus proton angle and missing energy for the
kinematical conditions considered in Figs. \protect
\ref{fig:centralx1} (left panel) and \protect \ref{fig:centralx2}
(right panel).}
\label{fig:superratio}
\end{figure}

\begin{figure}
\begin{center}
{\mbox{\epsfysize=8.cm\epsffile{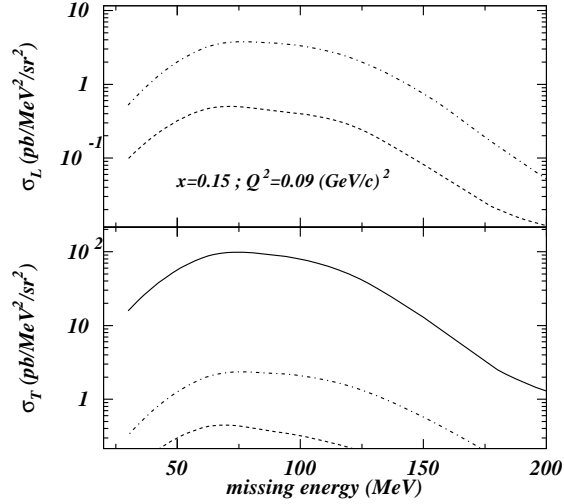}}}
\end{center}
\caption{The longitudinal ($\sigma _L$) and transverse ($\sigma _T$)
part of the differential $^{16}$O$(e,e'p)$ cross section versus
missing energy in parallel kinematics at $\epsilon$=1.2~GeV, $\epsilon
'$=0.9 ~GeV and $\theta_e$=16$^o$ (or, $x \approx 0.15 $ and $q =
0.42$~GeV/c).  The dashed line includes solely the central
correlations ; the dot-dashed line both central and tensor
correlations.  The solid curves represent the calculations that
include central and tensor correlations, as well as MEC and IC.
Neither the MEC nor the IC do contribute to the longitudinal part of
the cross section.}
\label{fig:parlowx}
\end{figure}

\begin{figure}
\begin{center}
{\mbox{\epsfysize=8.cm\epsffile{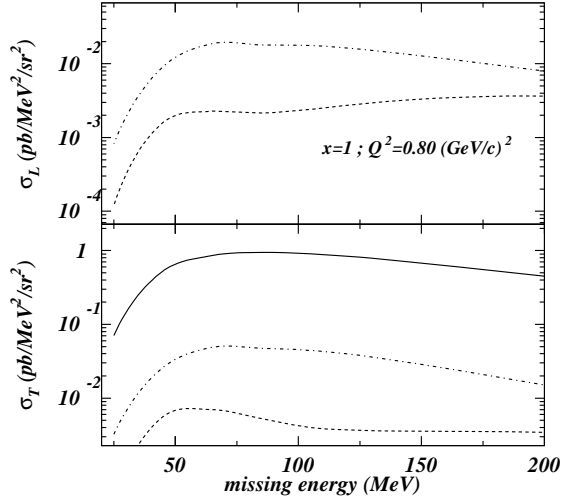}}}
\end{center}
\caption{As in Figure \protect \ref{fig:parlowx} but for
$\epsilon$=2.442~GeV, $\epsilon '$=1.997 ~GeV and $\theta_e$=23.4$^o$
(or, $x \approx 1$ and $q = 1.0$~GeV/c).}
\label{fig:parx1}
\end{figure}

\begin{figure}
\begin{center}
{\mbox{\epsfysize=8.cm\epsffile{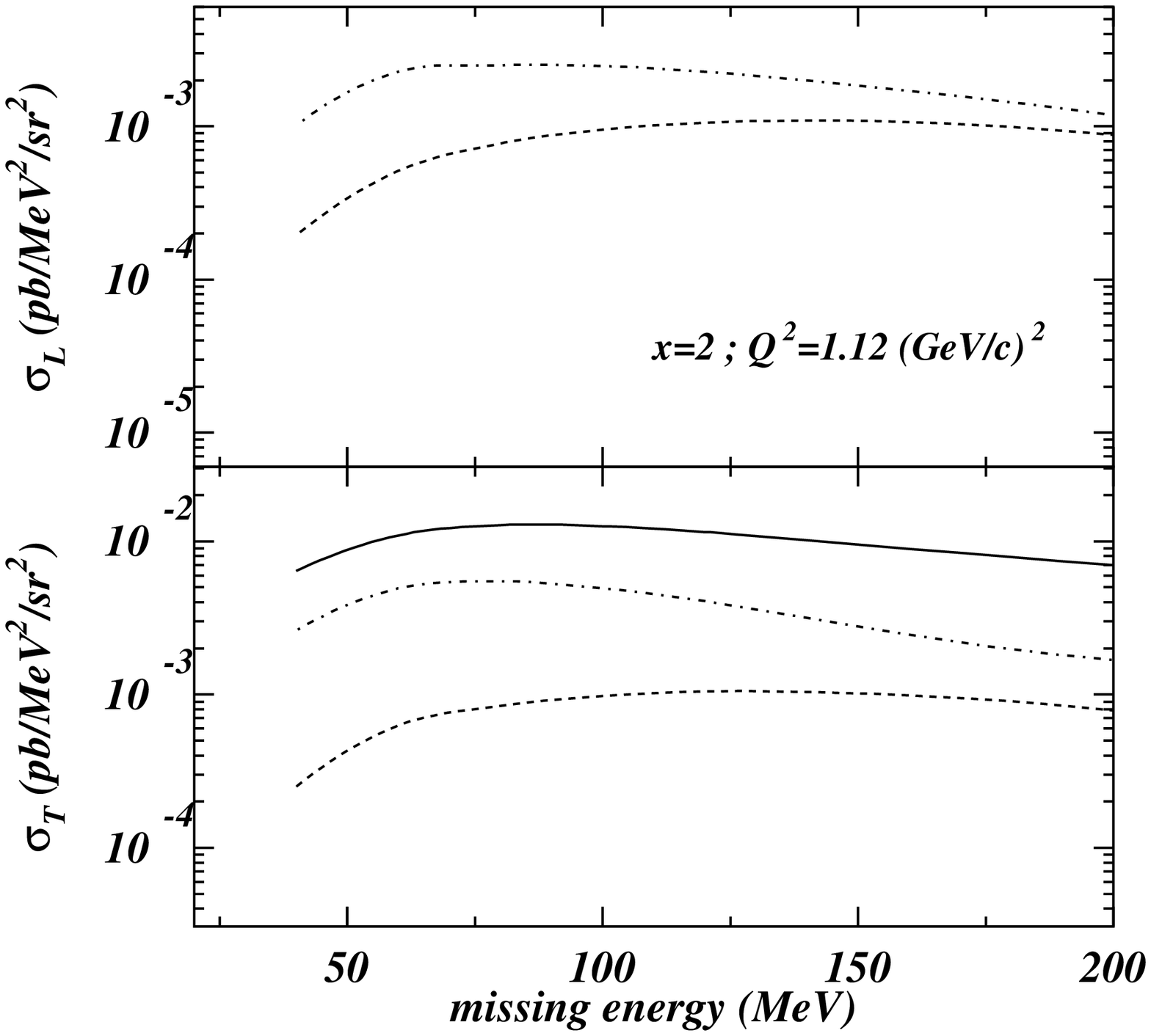}}}
\end{center}
\caption{As in Figure \protect \ref{fig:parlowx} but for
$\epsilon$=2.5~GeV, $\epsilon '$=2.2 ~GeV and $\theta_e$=26.0$^o$
(or, $x \approx 2$ and $q = 1.1$~GeV/c).}
\label{fig:parx2}
\end{figure}

A powerful tool in studies with electromagnetic probes is the
separation of the various structure functions.  We study the
longitudinal and transverse structure functions in parallel kinematics
($\theta _p = 0^o$) for all three kinematical situations which were
discussed earlier in this Sections.  Not only do solely two structure
functions ($W_L$ and $W_T$) contribute to the cross section in the
$\theta _p = 0^o$ case, parallel kinematics has often been shown to
create favorable conditions when it comes to controlling the fsi
effects \cite{frankfurt97,templon99,bianci95}. In high-energy
processes, for example, the eikonal approximation (or its multiple
scattering extension, usually referred to as Glauber theory) is
expected to be most accurate at small proton angles $\theta _{p}$.
One could wonder whether parallel kinematics is also favorable when it
comes to suppressing the background of MEC and IC in $(e,e'p)$.  To
that purpose we have evaluated the ratio of the calculated
$^{16}$O$(e,e'p)$ differential cross section with and without MEC/IC.
The results are shown in Fig.~\ref{fig:superratio}.  They illustrate
that parallel kinematics is not a bad choice with the eye on
suppressing the contribution from the genuine two-body currents (MEC
and IC).  In the $x \approx 2$ case the $NN$ correlations and two-body
currents are each responsible for about half of the calculated
strength at forward proton angles.  As one moves to higher proton
angles the $NN$ correlations gradually lose in importance.  For the $x
\approx 1$ case the $NN$ correlations account for a mere 5\% in
parallel kinematics and at $\theta _p$=90$^o$ their predicted impact
is virtually non-existent in comparison with the contributions from
the two-body currents.  The longitudinal and transverse part in the
differential cross sections for parallel kinematics are shown in
Figs.~\ref{fig:parlowx}-\ref{fig:parx2}.  For all three kinematical
conditions considered, the central and tensor $NN$ correlations
exhibit a similar qualitative behaviour in the longitudinal and
transverse part of the differential cross section.  For both structure
functions, the tensor term strongly dominates the strength attributed
to $NN$ correlations.  For $x \lesssim 1$ the central correlations
produce only small effects.  For higher values of x, when higher
missing momenta are probed, the central correlations gain in relative
importance with respect to the tensor correlations.  The transverse
response, though, is dramatically increased after including the
meson-exchange and isobar currents.  Qualitatively, the resulting missing
energy dependence (solid curves) of the transverse cross section
differs from the variant without two-body currents (dot-dashed curves)
insignificantly.

\section{Conclusions}
\label{sec:conclusions}

In summary, we have reported unfactorized calculations for the
$^{16}$O$(e,e'p)$ differential cross section at deep missing energies
and discussed its sensitivity to $NN$ correlations.  The calculations
are performed in a complete unfactorized theoretical framework in
which both state-dependent and state-independent $NN$ correlations are
incorporated, as well as competing processes related to genuine
two-body photoabsorption mechanisms like meson-exchange and isobar
currents.  We have evaluated the effect of central, spin-isospin and
tensor $NN$ correlations.  Of these three terms, the tensor
correlations are by far the strongest source of $(e,e'p)$ strength.
Their contribution often overshoots the strength from the central
(Jastrow) correlations by almost an order of magnitude.  The effect of
the spin-isospin $NN$ correlations upon the semi-exclusive $A(e,e'p)$
differential cross sections was found to be practically negligible.
The results reveal further that meson-exchange and isobar two-body
currents are strongly feeding the semi-exclusive $A(e,e'p)$ channel at
deep missing energies.  At low values of Bjorken $x$ the genuine
two-body currents (MEC/IC) typically produce one or two orders of
magnitude more $(e,e'p)$ strength above the two-nucleon knockout
threshold than the $NN$ correlations do. Even under quasi-elastic
conditions and four-momentum transfers of the order $Q^2 \approx
0.8~$(GeV/c)$^2$, the meson-exchange and isobar currents generate
substantially more $A(e,e'p)$ strength at deep missing energies than
the $NN$ correlations do.  The $NN$ correlations are found to equally
feed the longitudinal and transverse part of the differential cross
sections.  Whenever the meson-exchange and isobar currents start
dominating, the continuous part of the $(e,e'p)$ strength becomes
almost exclusively transverse.  At ``quasi-deuteron'' like values of
the Bjorken variable ($x \approx 2 $) the effect of the two-body
currents is observed to be sufficiently suppressed to make a direct
link between the measured $(e,e'p)$ differential cross sections and
the spectral function feasible.  We close with remarking that our
investigations reveal that the pion degrees of freedom tend to
dominate the $(e,e'p)$ strength that is not of single-particle origin.
Indeed, both the tensor $NN$ correlations and the
meson-exchange/isobar currents which we find to dominante the
semi-exclusive proton knockout process, are intimately connected to
the pion-exchange term in the nucleon-nucleon force.  Further, from
our investigations it appears hard to gain access to central (Jastrow)
correlations with the aid of the $(e,e'p)$ reaction.  Triple
coincidence $A(e,e'pp)$ measurements offer better perspectives in this
respect.

\vspace{0.5cm}
{\bf Acknowledgment}  This work was supported by the Fund for Scientific
Research of Flanders under contract No. 4.0061.99 and the University
Research Council.

\appendix
\label{sec:appendix}
\section{Matrix elements for electromagnetic coupling to a correlated
nucleon pair}

In Section~\ref{sec:effectiveop} it was pointed out that the standard
(one-body) electromagnetic coupling of a virtual photon to a tensor
and spin-isospin correlated nucleon pair can be formally treated as
effective two-body current operators that are composed of a
``correlation operator'' and a one-body hadronic current.  Here, we
collect the expressions for the two-body matrixelements that emerge
when considering the electromagnetic coupling through the standard
one-body operator of the impulse approximation to a tensor and
spin-isospin correlated pair.  The expressions for the central
(state-independent) correlation term can be found in
Ref.~\cite{jan97}.  As we rely on a partial-wave technique to deal
with the high dimensionality of the three-fragment final state, the
longitudinal and transverse multipole components are separately
constructed.  We first consider the matrix elements related to the
tensor (Section~\ref{sec:tensor}) and thereafter the spin-isospin
(Section~\ref{sec:spiniso}) correlations.  In describing the matrix
elements, we use the shorthand notation $ a \equiv \left( n_a, l_a,
\frac{1}{2}, j_a \right)$. If an other set of quantumnumbers is
involved, they are given explicitly.

\subsection{Tensor correlations}
\label{sec:tensor}
\subsubsection{Longitudinal contribution}

The effective operator that accounts for the coupling of a longitudinally
polarized virtual photon to a tensor correlated pair reads
\begin{equation}
\rho_{t\tau}^{ \left[ 1 \right]} \left( i,j \right) \equiv \frac{q}{\omega}
J_{\lambda=0, t \tau}^{ \left[ 1 \right],in} \left( i,j \right) 
=  e  \biggl[ \delta \! \left( \vec{r} - \vec{r}_i 
\right)  \frac{1+ \tau_{z,i}}{2} + \delta \! \left( \vec{r} -
\vec{r}_j \right) \frac{1+ \tau_{z,j}}{2} \biggr] 
\ \widehat{t} \left( i,j \right) \;.
\end{equation}
We define the partial wave components of the tensor correlation
function $f^{p=6}(r_{ij})$
\begin{equation}
{\cal X} ^{t \tau} \!  \left( l_1, l_2, r_i, r_j \right) \equiv \int \! dq \int \! dr
\ q^2 r^2 j_2 \!\left(qr \right) f^{p=6} \! \left(r \right) j_{l_1} \!
\left(qr_i \right) j_{l_2} \! \left( qr_j \right) \; , 
\end{equation}
and remark that the multipole components $M_{JM}^{coul}$ corresponding
with the above operator can be written in the form
\begin{eqnarray}
M_{JM}^{coul} & & \left[ \rho_{t\tau}^{ \left[ 1 \right]} \left( i,j
\right) \right] =  
\sum_{l_1 l_2} \sum_{L} \sum_{J_3 J_4} \frac{ 4 \sqrt{6}}{ \sqrt
{ \pi}} e {\cal X} ^ {t \tau} \!  \left( l_1, l_2, r_i, r_j \right)
\widehat{l}_1 \widehat{l}_2 \widehat{L} \widehat{J}_3 \widehat{  J}_4
\langle l_1 \ 0 \ l_2 \ 0 \mid 2 \ 0 \rangle  
\left\{ \begin{array}{ccc}
1 & 1 & 2 \nonumber \\
l_1 & l_2 & J_3 \end{array} \right\}
i^{l_1 + l_2} \left( \vec{ \tau}_i \cdot \vec{ \tau}_j
\right)  \nonumber \\ 
 & & \times \Biggl\{ \frac{1+ \tau_{z,i}}{2}\ \widehat{l}_1
\left( \begin{array}{ccc}
L & J & l_1 \nonumber \\
0 & 0 & 0 \end{array} \right)
\left\{ \begin{array}{ccc}
L & J & l_1 \nonumber \\
J_3 & 1 & J_4 \end{array} \right\} 
\left( -1 \right)^{J+1}  
j_J \left( qr_i \right) \biggl[ \left[ Y_L \left( \Omega_i
\right) \otimes \vec{ \sigma}_i \right]_{J_4} \otimes \left[ Y_{l_2}
\left( \Omega_j \right) \otimes \vec{ \sigma}_j \right]_{J_3}
\biggr]_J^M \nonumber \\ 
 & & + \frac{1+ \tau_{z,j}}{2}\  \widehat{l}_2
\left( \begin{array}{ccc}
L & J & l_2  \\
0 & 0 & 0 \end{array} \right)
\left\{ \begin{array}{ccc}
L & J & l_2  \\
J_3 & 1 & J_4 \end{array} \right\} 
\left( -1 \right)^{J_3+J_4+1} 
j_J \left( qr_j \right) \biggl[ \left[ Y_{l_1} \left
( \Omega_i \right) \otimes \vec{ \sigma}_i \right]_{J_3} \otimes
\left[ Y_L \left( \Omega_j \right) \otimes \vec{ \sigma}_j
\right]_{J_4} \biggr]_J^M \Biggr\} \;. 
\end{eqnarray}

After straightforward algebraic manipulations, the reduced two-body
matrix element corresponding with this operator reads 
\begin{eqnarray}
\langle ab ; J_1 \parallel  & & M_{JM}^{coul} \left[ \rho_{t
\tau} ^ { \left[ 1 \right]} \left( 1,2 \right) \right] \parallel cd ; J_2
\rangle = \sum_{l_1 l_2} \sum_{L} \sum_{J_3 J_4} \int \! dr_1 \int \!
dr_2   \frac{ 4 \sqrt{6}}{ \sqrt{ \pi}} e {\cal X} ^ {t \tau} \!
\left( l_1, l_2, r_1,  r_2 \right) \nonumber \\
 & & \times \widehat{l}_1 \widehat{l}_2 \widehat{L} \widehat{J}_3
 \widehat{  J}_4 \widehat{J}_1 \widehat{J}_2 \widehat{J} \nonumber
\langle l_1 \ 0 \ l_2 \ 0 \mid 2 \ 0 \rangle 
\left\{ \begin{array}{ccc}
1 & 1 & 2 \nonumber \\
l_1 & l_2 & J_3 \end{array} \right\}
i^{l_1 + l_2}  \nonumber \\ 
 & & \times \Biggl\{ \delta_{ac,p}\  \widehat{l}_1
\left( \begin{array}{ccc}
L & J & l_1 \nonumber \\
0 & 0 & 0 \end{array} \right)
\left\{ \begin{array}{ccc}
L & J & l_1 \nonumber \\
J_3 & 1 & J_4 \end{array} \right\} 
\left\{ \begin{array}{ccc}
j_a & j_b & J_1 \nonumber \\
j_c & j_d & J_2 \nonumber \\
J_4 & J_3 & J \end{array} \right\} \left( -1 \right)^{J+1} \nonumber \\
 & & \times \langle \ a \parallel j_J \left( qr_1 \right)  \left[ Y_L \left(
\Omega_1 \right)
\otimes \vec{ \sigma}_1 \right]_{J_4}  \parallel c \ \rangle_{r_1} \langle \ b
\parallel \left[
Y_{l_2} \left( \Omega_2 \right) \otimes \vec{ \sigma}_2 \right]_{J_3} \parallel
d \ 
\rangle_{r_2} \nonumber \\
 & & + \delta_{bd,p}\  \widehat{l}_2 
\left( \begin{array}{ccc}
L & J & l_2 \nonumber \\
0 & 0 & 0 \end{array} \right)
\left\{ \begin{array}{ccc}
L & J & l_2 \nonumber \\
J_3 & 1 & J_4 \end{array} \right\} 
\left\{ \begin{array}{ccc}
j_a & j_b & J_1 \nonumber \\
j_c & j_d & J_2 \nonumber \\
J_3 & J_4 & J \end{array} \right\} \left( -1 \right)^{J_3+J_4+1}
 \nonumber \\ 
 & & \times \langle \ a \parallel   \left[ Y_{l_1} \left( \Omega_1
 \right) \otimes \vec{  \sigma}_1 \right]_{J_3}  \parallel c \
 \rangle_{r_1} \langle \ b \parallel j_J \left( qr_2 \right)  \left
 [ Y_L \left( \Omega_2 \right) \otimes \vec{ \sigma}_2 \right]_{J_4}
 \parallel d \ \rangle_{r_2}  \Biggr\}  \; ,
\end{eqnarray}
where the radial transition densities $\langle \ a \parallel
\widehat{O} \parallel b \ \rangle_r $ are defined as
\begin{equation}
 \langle \ a \parallel
\widehat{O} \parallel b \ \rangle = \int \! dr \langle \ a \parallel
\widehat{O} \parallel b \ \rangle_r \;.  
\end{equation}

\subsubsection{Transverse contribution}
We introduce 
\begin{equation}
O_{JM}^{\kappa =0,\pm1} \left(q\right) = \sum_{M_1,M_2} \int d
\vec{r} \ \langle J\!\!+\!\!\kappa \ M_1 \ 1 \ M_2 \mid J M \rangle
Y_{J+\kappa M_1} \left( \Omega \right) j_{J+\kappa} \left ( qr
\right) J_{M_2} \left( \vec{r} \right) \; ,
\end{equation}
and remark that it suffices to determine the matrix elements of this
spherical tensor operator as both the electric and magnetic multipole
operators can be expressed in terms of $O_{JM}^{\kappa}$ 
\begin{eqnarray}
T_{JM}^{mag} \left(q\right) & \equiv & O_{JM}^{\kappa=0} \left(q\right)
\label{eq:TmagifvO} \; , \\
T_{JM}^{el} \left(q\right) & \equiv & \sum_{\kappa=\pm 1}   \frac{i  \left(-1
\right)^{ \delta_{  \kappa,1}}}{ \widehat{J} } \sqrt{ J + \delta_{\kappa, -1}}
\  O_{JM}^{\kappa} \left(q\right) \; .
\end{eqnarray}
In what follows, we treat the convective and
magnetization component of the transverse one-body current separately.
Electromagnetic coupling through the magnetization ($magn$) current to a
tensor-correlated pair 
\begin{equation}
\vec{J}_{t\tau} ^{ \ \left[  1 \right],magn} \left( i,j \right) = 
\frac{ e \hbar}{ 2 M_N} \left[ \mu_i \delta  \left( \vec{r} -
\vec{r}_i \right) \vec{  \nabla}_i \times \vec{ \sigma}_i 
+  \mu_j \delta \! \left( \vec{r} - \vec{r}_j \right) \vec
{  \nabla}_j \times \vec{ \sigma}_j \right] \  \widehat{t} \left( i,j \right)
\end{equation} 
leads to the following $O_{JM}^{\kappa}$ operator
\begin{eqnarray}
O_{JM}^{\kappa} & & \left[ \vec{J}_{t\tau} ^{ \ \left[  1 \right],magn}
\left( i,j \right) \right] =  
\sum_{l_1 l_2} \sum_{J_3 J_4} \sum_{L L_3} \sum_{ \eta= \pm1}
\frac{24}{ \sqrt{ \pi}} \frac{ \hbar eq}{2 M_N} \widehat{ l}_1
\widehat{ l}_2 \widehat{J}_3 \widehat{J}_4 \widehat{L} \widehat{L}_3
\widehat{ J\!\!+\!\! \kappa \!\!+\!\! \eta} \left( \vec{ \tau}_i \cdot
\vec{ \tau}_j \right) 
i^{l_1+l_2+1} 
\nonumber \\ & & \times \sqrt{J+ \kappa+ \delta_
{ \eta,+1}} {\cal X} ^ {t \tau} \!  \left( l_1, l_2, r_i, r_j \right)
\langle l_1 
\ 0 \ l_2 \ 0 \mid 2 \ 0 \rangle 
\left\{ \begin{array}{ccc}
1 & 1 & 2 \nonumber \\
l_2 & l_1 & J_3 \end{array} \right\}
\left\{ \begin{array}{ccc}
1 & J & J\!\!+\!\! \kappa \nonumber \\
J\!\!+\!\! \kappa \!\!+\!\! \eta & 1 & 1 \end{array} \right\}
\left\{ \begin{array}{ccc}
J & 1 & J\!\!+\!\! \kappa \!\!+\!\! \eta \nonumber \\
J_4 & J_3 & L_3 \end{array} \right\}
\nonumber \\
 & & \times \Biggl\{ \left[ \delta_{i,p} \mu_p + \delta_{i,n}
\mu_n \right] \widehat{l}_1 
\left( \begin{array}{ccc}
L & l_1 & J\!\!+\!\! \kappa \!\!+\!\! \eta \nonumber \\
0 & 0 & 0 \end{array} \right)
\left\{ \begin{array}{ccc}
L &  J\!\!+\!\! \kappa \!\!+\!\! \eta & l_1 \nonumber \\
J_3 & 1 & J_4 \end{array} \right\}  \left(-1 \right)^{ L_3 - J_3 +
\kappa} \nonumber \\ 
 & & \times j_{J + \kappa + \eta} \left(qr_i \right) \Biggl[ \biggl
[ \left[ Y_L \left( \Omega_i \right) \otimes \vec{ \sigma}_i
\right]_{J_4} \otimes \vec{ \sigma}_i \biggr]_{L_3} \otimes \left[
Y_{l_2} \left( \Omega_j \right) \otimes \vec{ \sigma}_j \right]_{J_3}
\Biggr]_J^M  \nonumber \\
 & & + \left[ \delta_{j,p} \mu_p + \delta_{j,n} \mu_n \right]
\widehat{l}_2 
\left( \begin{array}{ccc}
L & l_2 & J\!\!+\!\! \kappa \!\!+\!\! \eta \nonumber \\
0 & 0 & 0 \end{array} \right)
\left\{ \begin{array}{ccc}
L &  J\!\!+\!\! \kappa \!\!+\!\! \eta & l_2 \nonumber \\
J_3 & 1 & J_4 \end{array} \right\}  
\left(-1 \right)^{ J + \kappa} \nonumber \\
 & & \times j_{J + \kappa + \eta} \left(qr_j \right) \Biggl[ \left
[ Y_{l_1} \left( \Omega_i \right)  \otimes \vec{ \sigma}_i
\right]_{J_3} \otimes \biggl[ \left[ Y_L \left( \Omega_j \right)
\otimes \vec{ \sigma}_j \right]_{J_4} \otimes \vec{ \sigma}_j
\biggr]_{L_3}  \Biggr]_J^M \Biggr\} \; . 
\end{eqnarray}
With the aid of standard Racah algebra techniques it can be shown that the
reduced matrix element corresponding with this operator can be cast in
the form
\begin{eqnarray}
 \langle ab; J_1 & \parallel & O_{J}^{\kappa} \left[ \vec{J}_{t \tau}^
{ \  \left[ 1 \right],magn} \left( 1,2 \right) \right] \parallel cd;
J_2  \rangle =  
\sum_{l_1 l_2} \sum_{J_3 J_4} \sum_{L L_3} \sum_{ \eta= \pm1}
\sum_j  {\int \! dr_1 \int \!  dr_2} \frac{24 \sqrt{6}}{ \sqrt{ \pi}}
\frac{ \hbar eq}{2 M_N} \langle l_1 \ 0 \ l_2 \ 0 \mid 2 \ 0 \rangle 
\nonumber \\  & & 
\times \widehat{ l}_1 \widehat{ l}_2 \widehat{J}_3 \widehat{J}_4
\widehat{L} \left( \widehat{L}_3 \right)^2 \widehat{ J\!\!+\!\! \kappa
\!\!+\!\! \eta} \widehat{J}_1 \widehat{J}_2 \widehat{J}  \widehat{j}
i^{l_1+l_2+1} \sqrt{J+ \kappa+ \delta_{ \eta,+1}} {{\cal
X}^{t \tau} \! \left( l_1, l_2, r_1, r_2 \right)} \nonumber \\ & & \times
\left\{ \begin{array}{ccc}
1 & 1 & 2 \nonumber \\ 
l_2 & l_1 & J_3 \end{array} \right\} 
\left\{ \begin{array}{ccc} 
1 & J & J\!\!+\!\! \kappa \nonumber \\ 
J\!\!+\!\! \kappa \!\!+\!\! \eta & 1 & 1 \end{array} \right\} 
\left\{ \begin{array}{ccc}
J & 1 & J\!\!+\!\! \kappa \!\!+\!\! \eta \nonumber \\
J_4 & J_3 & L_3 \end{array} \right\} \nonumber \\
& & \times \Biggl\{ \left[ \delta_{ac,p} \mu_p + \delta_{ac,n}
\mu_n \right]   \widehat{l}_1 \widehat{j}_c
\left( \begin{array}{ccc}
L & l_1 & J\!\!+\!\! \kappa \!\!+\!\! \eta \nonumber \\
0 & 0 & 0 \end{array} \right)
\left\{ \begin{array}{ccc}
L &  J\!\!+\!\! \kappa \!\!+\!\! \eta & l_1 \nonumber \\
J_3 & 1 & J_4 \end{array} \right\}
\nonumber \\
 & & \left\{ \begin{array}{ccc}
J_4 & 1 & L_3 \nonumber \\
j_c & j_a & j \end{array} \right\}
\left\{ \begin{array}{ccc}
l_c & 1/2 & j \nonumber \\
 1 & j_c & 1/2 \end{array} \right\}
\left\{ \begin{array}{ccc}
j_a & j_b & J_1 \nonumber \\ 
j_c & j_d & J_2 \nonumber \\ 
L_3 & J_3 & J \end{array} \right\}
\left(-1 \right)^{ j_a + j_c + l_c + j - J_3 + \kappa + 3/2} \nonumber\\ 
 & & \times \langle \ a \parallel j_{J+ \kappa+ \eta} \left( qr_1
\right) \left[ Y_L \left( \Omega_1 \right) \otimes \vec{ \sigma}_1
\right]_{J_4} \  \parallel n_c l_c \frac{1}{2} j \ \rangle_{r_1}
\langle \ b \parallel  \left[ Y_{l_2} \left( \Omega_2 \right)  \otimes
\vec{ \sigma}_2  \right]_{J_3} \parallel d \ \rangle_{r_2}  \nonumber \\
 & & + \left[ \delta_{bd,p} \mu_p + \delta_{bd,n} \mu_n \right]
\widehat{l}_2  \widehat{j}_d 
\left( \begin{array}{ccc}
L & l_2 & J\!\!+\!\! \kappa \!\!+\!\! \eta \nonumber \\
0 & 0 & 0 \end{array} \right)
\left\{ \begin{array}{ccc}
L &  J\!\!+\!\! \kappa \!\!+\!\! \eta & l_2 \nonumber \\
J_3 & 1 & J_4 \end{array} \right\}
\nonumber \\
 & & 
\left\{ \begin{array}{ccc}
J_4 & 1 & L_3 \nonumber \\
j_d & j_b & j \end{array} \right\}
\left\{ \begin{array}{ccc}
l_d & 1/2 & j \nonumber \\
 1  & j_d & 1/2 \end{array} \right\}
\left\{ \begin{array}{ccc}
j_a & j_b & J_1 \nonumber \\
j_c & j_d & J_2 \nonumber \\
J_3 & L_3 & J \end{array} \right\} 
\left(-1 \right)^{ j_b + j_d + l_d + j + L_3 + J + \kappa + 3/2} \nonumber \\
 & & \times \langle \ a \parallel \left[ Y_{l_1} \left( \Omega_1
\right)  \otimes \vec{ \sigma}_1 \right]_{J_3} \parallel c \
\rangle_{r_1} \langle \ b \parallel  j_{J+ \kappa+ \eta} \left( qr_2
\right) \left[ Y_L \left( \Omega_2 \right) \otimes \vec{ \sigma}_2
\right]_{J_4}  \parallel n_d l_d \frac{1}{2} j \ \rangle_{r_2}
\Biggr\} \;.  
\end{eqnarray}

The combination of the electromagnetic coupling through the convection
current ($conv$) with a tensor correlated pair 
\begin{equation}
\vec{J}_{t\tau} ^{ \ \left[ 1 \right],conv} \left( i,j \right) = 
\frac{ e \hbar}{ 2 M_N i}  \left[ \frac{1+ \tau_{z,i}}{2} 
\left\{ \vec{\nabla}_i , \delta  \left( \vec{r} - \vec{r}_i \right)
\right\} 
+ \frac{1+ \tau_{z,j}}{2} \left\{ \vec{ \nabla}_j,  \delta 
\left( \vec{r} - \vec{r}_j \right) \right\} \right] \  \widehat{t}
\left(i,j \right)   
\end{equation}
leads to the following spherical tensor operator
\begin{eqnarray}
O_{JM}^{\kappa} & & \left[ \vec{J}_{t\tau} ^{ \ \left[ 1 \right],conv}
\left( i,j \right) \right] =  
\sum_{l_1 l_2}
\sum_{J_3 J_4} \sum_{J_5 L} \sqrt{\frac{24}{\pi}} \frac{ \hbar e}{M_N}
\widehat{ l}_1 \widehat{ l}_2 \widehat{J}_3 \widehat{J}_4
\widehat{J}_5 \widehat{L} \widehat{ J\!\!+\!\! \kappa} \left( \vec{
\tau}_i \cdot \vec{ \tau}_j \right) \nonumber \\ 
& & \times i^{l_1+l_2-1} {\cal X} ^ {t \tau} \!  \left(l_1, l_2, r_i,
r_j \right) 
\langle l_1 \ 0 \ l_2 \ 0 \mid 2 \ 0 
\rangle \left\{ \begin{array}{ccc} 
1 & 1 & 2 \nonumber \\ 
l_2 & l_1 & J_3 \end{array} \right\} \nonumber \\ 
& & \times \Biggl\{ \left( \frac{1 + \tau_{z,i}}{2} \right)
\widehat{l}_1 
\left( \begin{array}{ccc} L & l_1 & J\!\!+\!\! \kappa \nonumber \\ 
0 & 0 & 0 \end{array} \right) 
\left\{ \begin{array}{ccc} 
1 & J & J\!\!+\!\! \kappa \nonumber \\ 
l_1 & L & J_4 \end{array} \right\} 
\left\{ \begin{array}{ccc} 
1 & J_3 & l_1 \nonumber \\ 
J & J_4 & J_5 \end{array} \right\} 
\left( -1 \right)^{J} \nonumber \\ 
& & \times j_{J + \kappa} \left(qr_i \right) \Biggl[ \biggl[ \left
[ Y_L \left( \Omega_i \right) \otimes \left( \vec{\nabla}_i -
\vec{\nabla}_i' \right) \right]_{J_4} \otimes \vec{ \sigma}_i
\biggr]_{J_5} \otimes \left[ Y_{l_2} \left( \Omega_j \right) \otimes
\vec{ \sigma}_j \right]_{J_3} \Biggr]_J^M \nonumber \\  
& & + \left( \frac{1 + \tau_{z,j}}{2} \right) \widehat{l}_2 
\left( \begin{array}{ccc} 
L & l_2 & J\!\!+\!\! \kappa \nonumber \\ 
0 & 0 & 0 \end{array} \right) 
\left\{ \begin{array}{ccc} 
1 & J & J\!\!+\!\! \kappa \nonumber \\ 
l_2 & L & J_4 \end{array} \right\} 
\left\{ \begin{array}{ccc} 
1 & J_3 & l_2 \nonumber \\ 
J & J_4 & J_5 \end{array} \right\} 
\left(-1 \right)^{J_3 + J_5} \nonumber \\ 
& & \times j_{J + \kappa} \left(qr_j \right) \Biggl[ \left[ Y_{l_1}
\left( \Omega_i \right) \otimes \vec{ \sigma}_i \right]_{J_3} \otimes
\biggl[ \left[ Y_L \left( \Omega_j \right) \otimes \left
( \vec{\nabla}_j - \vec{\nabla}_j' \right) \right]_{J_4} \otimes \vec
{ \sigma}_j  \biggr]_{J_5} \Biggr]_J^M \Biggr\} \; .
\end{eqnarray}
In deriving the above expressions, the terms containing derivatives of
the correlation function are neglected. The notation $\vec{\nabla}_j'$
denotes a gradient operator acting to the left.  The reduced
matrix element corresponding with the above operator reads
\begin{eqnarray}
\langle ab; J_1 \parallel O_{J}^{\kappa} & & \left[
\vec{J}_{t\tau}^{ \ \left[ 1 \right],conv} \left( 1,2 \right) \right]
\parallel cd; J_2 \rangle =  \sum_{l_1 l_2}
\sum_{J_3 J_4} \sum_{J_5 L} \int \! dr_1 \int \!  dr_2 \frac{12}{
\sqrt{ \pi}} \frac{ \hbar e}{ M_N} \widehat{ l}_1 \widehat{ l}_2
\widehat{J}_1 \widehat{J}_2 \widehat{J}_3 \widehat{J}_4 \left(
\widehat{J}_5 \right)^2 \widehat{L} \widehat{ J\!\!+\!\! \kappa}
\widehat{J}  \nonumber \\ 
& & \times \langle l_1 \ 0 \ l_2 \ 0 \mid 2 \ 0 \rangle 
\left\{ \begin{array}{ccc} 
1 & 1 & 2 \nonumber \\ 
l_2 & l_1 & J_3 \end{array} \right\} 
i^{l_1+l_2-1} {\cal X} ^ {t \tau} \!  \left( l_1, l_2, r_1, r_2 \right)
\nonumber \\ 
& & \times \Biggl\{ \delta_{ac,p} \; \widehat{l}_1 \widehat{j}_a
\widehat{j}_c 
\left( \begin{array}{ccc} 
L & l_1 & J\!\!+\!\! \kappa \nonumber \\ 
0 & 0 & 0 \end{array} \right) 
\left\{ \begin{array}{ccc}
1 & J & J\!\!+\!\! \kappa \nonumber \\ 
l_1 & L & J_4 \end{array} \right\} 
\left\{ \begin{array}{ccc} 
1 & J_3 & l_1 \nonumber \\ 
J & J_4 & J_5 \end{array} \right\}
\left\{ \begin{array}{ccc} 
j_a & j_b & J_1 \nonumber \\ 
j_c & j_d & J_2 \nonumber \\ 
J_5 & J_3 & J \end{array} \right\} 
\left\{ \begin{array}{ccc} 
l_a & 1/2 & j_a \nonumber \\ 
l_c & 1/2 & j_c \nonumber \\ 
J_4 & 1 & J_5 \end{array} \right\} 
\left(-1 \right)^{J} \nonumber \\ 
& & \times \langle \ n_a l_a \parallel j_{J+ \kappa} \left( qr_1 \right) 
\left[ Y_L \left( \Omega_1 \right) \otimes \left( \vec{ \nabla}_1 - 
\vec{ \nabla}_1' \right) \right]_{J_4} \parallel n_c l_c  \rangle_{r_1} 
\langle \ b \parallel \left[ Y_{l_2} \left( \Omega_2 \right) \otimes 
\vec{ \sigma}_2 \right]_{J_3} \parallel d \ \rangle_{r_2} 
\nonumber \\ 
& & + \delta_{bd,p} \; \widehat{l}_2 \widehat{j}_b \widehat{j}_d 
\left(  \begin{array}{ccc}
L & l_2 & J\!\!+\!\! \kappa \nonumber \\ 
0 & 0 & 0 \end{array} \right)
\left\{ \begin{array}{ccc} 
1 & J & J\!\!+\!\! \kappa \nonumber \\ 
l_2 & L & J_4 \end{array} \right\}  
\left\{ \begin{array}{ccc} 
1 & J_3 & l_2 \nonumber \\ 
J & J_4 & J_5 \end{array} \right\}
\left\{ \begin{array}{ccc} 
j_a & j_b & J_1 \nonumber \\ 
j_c & j_d & J_2 \nonumber \\ 
J_3 & J_5 & J \end{array} \right\}
\left\{ \begin{array}{ccc} 
l_b & 1/2 & j_b \nonumber \\ 
l_d & 1/2 & j_d \nonumber \\ 
J_4 & 1 & J_5 \end{array} \right\}  
\left(-1 \right)^{J_3 + J_5} \nonumber \\ 
& & \times \langle \ a \parallel \left[ Y_{l_1} \left( \Omega_1 
\right) \otimes \vec{ \sigma}_1 \right]_{J_3} \parallel c \ 
\rangle_{r_1} \langle \ n_b l_b \parallel j_{J+ \kappa} \left( qr_2 \right) 
\left[ Y_L \left( \Omega_2 \right) \otimes \left( \vec{ \nabla}_2 - 
\vec{ \nabla}_2' \right) \right]_{J_4} \parallel n_d l_d \ \rangle_{r_2}
\Biggr\} \;.
\end{eqnarray}

\subsection{Spin-isospin correlations}
\label{sec:spiniso}
\subsubsection{Longitudinal contribution}

In a completely analogous manner as outlined for the tensor
correlations in Section~\ref{sec:tensor}, the effective two-body
operators and matrix elements corresponding with the coupling of a
virtual photon to a spin-isospin correlated dinucleon, i.e. a nuclear
pair correlated through the operator $f^{p=4} \left( r_{ij} \right)
\left( \vec{ \sigma}_i \cdot \vec { \sigma}_j \right) \left( \vec{
\tau}_i \cdot \vec{ \tau}_j \right) $, can be constructed.  The
effective Coulomb operator of rank $JM$ and corresponding reduced
two-body matrix element for longitudinal electromagnetic coupling to a
spin-isospin correlated pair becomes
\begin{eqnarray}
M_{JM}^{coul} \left[ \rho_{\sigma \tau}^{ \left[ 1 \right]} \left( i,j
\right)  \right] & = &
\sum_{lL} \sum_{J_3 J_4} \sqrt{ 4 \pi}e \frac{ \widehat{L}
\widehat{J}_3 \widehat{J}_4}{ \widehat{l}} \ {\cal X}^{\sigma \tau}
\left(l,  r_i, 
r_j  \right) \left( \vec{ \tau}_i \cdot \vec{ \tau}_j \right)
\left( \begin{array}{ccc}
J & L & l \nonumber \\
0 & 0 & 0
\end{array} \right)  
\left\{ \begin{array}{ccc}
J_3 & L & 1 \nonumber \\
l & J_4 & J
\end{array} \right\}  
\nonumber \\
 & & \times \Biggl\{ \left( -1 \right)^{J_4+L} j_J \left( q r_i
\right) \Biggl[ \biggl[ Y_L \left( \Omega_i \right)
\otimes  \vec{ \sigma}_i \biggr]_{J_3} \otimes \biggl[ Y_l \left
( \Omega_j \right) \otimes \vec{ \sigma}_j \biggr]_{J_4}  \Biggr]_J^M
\frac{1+ \tau_{z,i}}{2} \nonumber \\
 & & + \left( -1 \right)^{J_3+L+J} j_J \left( q r_j
\right) \Biggl[ \biggl[ Y_l \left( \Omega_i \right)
\otimes  \vec{ \sigma}_i \biggr]_{J_4} \otimes \biggl[ Y_L \left
( \Omega_j  \right) \otimes \vec{ \sigma}_j \biggr]_{J_3}  \Biggr]_J^M
\frac{1+ \tau_{z,j}}{2}  \Biggr\} \;,
\end{eqnarray}

\begin{eqnarray}
\langle ab; J_1 & \parallel & M_{J}^{coul} \left[ \rho_{\sigma \tau}^{ 
\left[ 1  \right]} \left( 1,2 \right) \right] \parallel cd; J_2
\rangle =  \sum_{lL} \sum_{J_3 J_4} \int \! d r_1 \int \! d r_2  \sqrt{ 4
\pi}e  \frac{ \widehat{ J}_1 \widehat{ J}_2 \widehat{ J}_3 
\widehat{ J}_4 \widehat{L}  }{ \widehat{l}} \langle l \ 0 \ L \ 0  \mid J \ 0
\rangle \  
\left\{ \begin{array}{ccc}
J_3 & L & 1 \nonumber \\ 
l & J_4 & J
\end{array} \right\}
{\cal X}^{\sigma \tau} \left( l, r_1, r_2 \right)
\nonumber \\
 & & \times \Biggl[ \delta_{ac,p} \left( -1 \right)^{l + J_4}
\left\{ \begin{array}{ccc}
j_a & j_b & J_1 \nonumber \\
j_c & j_d & J_2 \nonumber \\
J_3 & J_4 & J   \end{array} \right\}
 \langle \ a \parallel j_J \left( q r_1 \right) \biggl[ Y_L
 \left ( \Omega_1 \right) \otimes \vec{ \sigma}_1 \biggr]_{J_3}
 \parallel c \ \rangle_{r_1} \langle \ b \parallel \biggl[ Y_l \left(
 \Omega_2 \right) \otimes \vec{ \sigma}_2 \biggr]_{J_4}
\parallel d \ \rangle_{r_2}  
\nonumber \\
 & & + \delta_{bd,p} \left( -1 \right)^{l+J+J_3}
\left\{ \begin{array}{ccc}
j_a & j_b & J_1  \\
j_c & j_d & J_2  \\
J_4 & J_3 & J   \end{array} \right\} 
\langle \ a  \parallel  \biggl[ Y_l \left( \Omega_1
\right)  \otimes \vec{ \sigma}_1 \biggr]_{J_4} \parallel c \
\rangle_{r_1} \langle  \ b \parallel \biggl[ j_J \left( q r_2 \right)
Y_L  \left( \Omega_2 \right) \otimes \vec{ \sigma}_2 \biggr]_{J_3}
\parallel d \ \rangle_{r_2} 
\Biggr] \;.
\end{eqnarray}
The partial wave components of the spin-isospin correlation function
$f^{p=4} \left( r_{ij} \right)$ which appear in the above expressions take the form
\begin{equation}
{\cal X}^{\sigma \tau} \left(l, r_i, r_j \right) = \frac{2l+1}{2}
\int^{+1}_{-1} dx \ P_l \left(x \right) f^{p=4} \left( \sqrt{ r_i^2 +
r_j^2 - 2\ r_i r_j x} \right) \;.
\end{equation}

\subsubsection{Transverse contribution}
The combination of the transverse magnetization current and the
spin-isospin correlation operator $\left( f^{p=4} \left( r_{ij}
\right) \left( \vec{  \sigma}_i \cdot \vec{ \sigma}_j \right) \left
( \vec{ \tau}_i \cdot \vec{ \tau}_j \right) \right)$ leads to the
following transverse multipole operator  
\begin{eqnarray}
O_{JM}^{\kappa} & & \left[ \vec{J}_{\sigma \tau}^{ \ \left[ 1 \right],magn}
\left(  i,j \right) \right] =   
\sum_{lL} \sum_{J_3 J_4} \sum_{J_5 J_6}  \sum_{ \eta= \pm1}
 \frac{ i e \hbar q}{2 M_N} \sqrt{24 \pi}  \frac{ \widehat{L}
 \widehat{J}_3 \widehat{J}_4 (\widehat{J}_5)^2 \widehat{J}_6 }
 { \widehat{l}}\ {\cal X}^{\sigma \tau} \left(l, r_i, r_j \right) \sqrt{ J
+ \kappa + 
\delta_{ \eta, +1} } \nonumber \\
 & & \times  
\langle l \ 0 \ L \ 0 \mid J\!\!+\!\! \kappa \!\!+ \!\! \eta  \ 0  \rangle 
\left\{ \begin{array}{ccc}
1 & 1 & 1 \nonumber \\
J\!\!+\!\! \kappa & J\!\!+\!\! \kappa\!\! + \!\!\eta & J \end{array}  \right\}
\left\{ \begin{array}{ccc}
l & L  & J\!\!+\!\! \kappa\!\! + \!\!\eta  \nonumber \\
1 & J & J_5  \end{array} \right\}
\left\{ \begin{array}{ccc}
J_5 & 1  & L \nonumber \\
J_3 & 1 & J_6  \end{array} \right\}
\left\{ \begin{array}{ccc}
J_6 & J_5  & 1 \nonumber \\
l & J_4 & J  \end{array} \right\}
 \nonumber \\
& & \times \Biggl\{ \left[ \mu_p \delta_{i,p} + \mu_n
\delta_{i,n} \right] \Biggl[ \biggl[ \left[ Y_L \left( \Omega_i
\right)  \otimes  \vec{ \sigma}_i \right]_{J_3}  \otimes  
\vec{ \sigma}_i \biggr]_{J_6} \otimes \left[ Y_l \left( \Omega_j \right)
\otimes  \vec{ \sigma}_j \right]_{J_4} \Biggr]_J^M
j_{ J+\kappa + \eta } \left( qr_i \right) \left( -1 \right)^
 {L+J_3+J_4+J_5+J+\kappa+1 } \nonumber \\
 & & + \left[ \mu_p \delta_{j,p} + \mu_n
\delta_{j,n} \right] \Biggl[ \left[ Y_l \left( \Omega_i
\right)  \otimes  \vec{ \sigma}_i \right]_{J_4} \otimes 
\biggl[  \left[ Y_L  \left(  \Omega_j \right) \otimes  \vec{ \sigma}_j
\right]_{J_3} \otimes \vec{  \sigma}_j \biggr]_{J_6} \Biggr]_J^M
j_{ J+\kappa + \eta } \left( qr_j \right) \left( -1 \right)^
 { L+J_3+J_5+J_6+\kappa+1}
\Biggr\} \; .
\end{eqnarray}
The reduced matrix element of this multipole operator becomes: 
\begin{eqnarray}
\langle ab; J_1  & \parallel &  O_{J}^{\kappa} \left
[ \vec{J}_{\sigma \tau }^{\  \left[ 1 \right],magn} \left( 1,2 \right)
 \right] \parallel cd;  J_2 \rangle = 
 \sum_{lL}  \sum_{J_3 J_4} \sum_{J_5 J_6} \sum_{j}  \sum_{ \eta =
 \pm1} \int \! d r_1 \int \! d r_2 \frac{ i e \hbar q}{2 M_N} 12
 \sqrt{ \pi} \frac{ \widehat{L} \widehat{J}_1 \widehat{J}_2
 \widehat{J}_3  \widehat{J}_4 (\widehat{J}_5)^2 (\widehat{J}_6)^2
 \widehat{J} \widehat{j} }{ \widehat{l}} {\cal X}^{\sigma \tau} \left( l , r_1, r_2 \right)  
\nonumber \\
 & & \times 
\sqrt{ J \!+ \! \kappa \!+ \! \delta_{ \eta, +1} } \  
\langle l \ 0 \ L \ 0 \mid J\!\!+\!\! \kappa \!\!+ \!\! \eta  \ 0  \rangle 
\left\{ \begin{array}{ccc}
1 & 1 & 1 \nonumber \\
J\!\!+\!\! \kappa & J\!\!+\!\! \kappa\!\! + \!\!\eta & J \end{array}  \right\}
\left\{ \begin{array}{ccc}
l & L  & J\!\!+\!\! \kappa\!\! + \!\!\eta  \nonumber \\
1 & J & J_5  \end{array} \right\}
\left\{ \begin{array}{ccc}
J_5 & 1  & L \nonumber \\
J_3 & 1 & J_6  \end{array} \right\}
\left\{ \begin{array}{ccc}
J_6 & J_5  & 1 \nonumber \\
l & J_4 & J  \end{array} \right\}
 \nonumber \\
 & & \times \Biggl[ \left[ \mu_p \delta_{ac,p} + \mu_n
\delta_{ac,n} \right] \widehat{j}_c
\left( -1 \right)^{L+J_3+J_4+J_5+J_6+J+\kappa+j+j_a+j_c+l_c+1/2} 
\nonumber \\
 & & \times
\left\{ \begin{array}{ccc}
J_3 & 1  & J_6 \nonumber \\
j_c & j_a & j  \end{array} \right\}
\left\{ \begin{array}{ccc}
l_c & 1/2  & j \nonumber \\
1 & j_c & 1/2  \end{array} \right\}
\left\{ \begin{array}{ccc}
j_a & j_b & J_1 \nonumber \\
j_c & j_d & J_2 \nonumber \\
J_6 & J_4 & J   \end{array} \right\}
 \nonumber \\  
 & & \times \langle \ a \parallel j_{J+ \kappa+ \eta} \left( qr_1
 \right) \left[  Y_L \left( \Omega_1 \right) \otimes \vec{ \sigma}_1
 \right]_{J_3} \parallel n_c l_c \frac{1}{2} j \ \rangle_{r_1}
\langle \ b \parallel \left[ Y_l \left( \Omega_2 \right) \otimes \vec
 { \sigma}_2 \right]_{J_4}  \parallel d \ \rangle_{r_2}  
\nonumber \\
\nonumber \\
 & & +\left[ \mu_{ \pi} \delta_{bd, \pi} + \mu_{ \nu}
\delta_{bd, \nu} \right] \widehat{j}_d
\left( -1 \right)^{L+J_3+J_5+\kappa+j+j_b+j_d+l_d+1/2} 
\nonumber \\ 
 & & \times
\left\{ \begin{array}{ccc}
J_3 & 1  & J_6 \nonumber \\
j_d & j_b & j  \end{array} \right\}
\left\{ \begin{array}{ccc}
l_d & 1/2  & j \nonumber \\
1 & j_d & 1/2  \end{array} \right\}
\left\{ \begin{array}{ccc}
j_a & j_b & J_1 \nonumber \\
j_c & j_d & J_2 \nonumber \\
J_4 & J_6 & J   \end{array} \right\}
 \nonumber \\ 
 & & \times \langle \ a \parallel  \left[  Y_l \left( \Omega_1 \right)
 \otimes \vec{ \sigma}_1  \right]_{J_4} \parallel c \ \rangle_{r_1}
\langle \ b \parallel j_{J+ \kappa+ \eta} \left( qr_2
 \right) \left[ Y_L \left( \Omega_2 \right) \otimes \vec
 { \sigma}_2 \right]_{J_3}  \parallel n_d l_d \frac{1}{2} j  \ \rangle_{r_2}  
 \Biggr] \;.
\end{eqnarray}

For both the central and the tensor correlation term the numerical
calculations indicate that the transverse strength is almost exclusivly due
to the magnetization current, the convection current producing very
small matrix elements.  For that reason we have neglected the convection
current contribution when evaluating the spin-spin correlations.

\begin{thebibliography}{99}
\bibitem{kester96}
L.J.H.M. Kester {\em et al.}, Phys. Lett. B366 (1996) 44.
\bibitem{frankfurt97}
L.L. Frankfurt, M.M. Sargsian and M.I. Strikman, Phys. Rev. C 56
(1997) 1124.
\bibitem{ingoelba}
I. Sick in Proceedings of the Workshop on Electron Nucleus Scattering,
eds Omar Benhar and Adelchi Fabrocini (Edizioni ETS, Pisa, 1997) p. 445.
\bibitem{demetriou99} P. Demetriou, A. Gil, S. Boffi, C. Giusti,
E. Oset and F.D. Pacati, Nucl. Phys. A650 (1999) 199.
\bibitem{amparo98} A. Gil, J. Nieves and E. Oset, Nucl. Phys. A627 (1997) 599.
\bibitem{benhar99}
O. Benhar, S. Fantoni and G.I. Lykasov, Eur. Phys. J A  5 (1999) 137.
\bibitem{morita99}
H. Morita, C. Ciofi degli Atti and D. Treleani, Phys. Rev. C 60
(1999) 034603. 
\bibitem{claudio99} Claudio Ciofi degli Atti and Daniele Treleani,
Phys. Rev. C 60 (1999), 024602.
\bibitem{ulmer} P.E. Ulmer {\em et al.}, Phys. Rev. Lett. 59
(1987) 2259.
\bibitem{lourie93} R.W. Lourie, W. Bertozzi, J. Morrison and
L.B. Weinstein, Phys. Rev. C 57 (1993) R444.
\bibitem{liyanage}
N. Liyanage, Ph.D. thesis, MIT (1999).
\bibitem{maurik}
M. Holtrop {\em et al.}, Phys. Rev. C 58 (1998) 3205.
\bibitem{dutta}
D.~Dutta, Ph.D. thesis, Northwestern University (1999).
\bibitem{jan97} J. Ryckebusch, V. Van der Sluys, K. Heyde, H. Holvoet,
W. Van Nespen, M. Waroquier and M. Vanderhaeghen, Nucl. Phys. A 624 (1997) 581.
\bibitem{jana568}
J. Ryckebusch, M. Vanderhaeghen, L. Machenil and M. Waroquier,
Nucl. Phys. A 568 (1994) 828-854. 
\bibitem{pieper} S.C. Pieper, R.B. Wiringa and V.R. Pandharipande,
Phys. Rev. C 46 (1992) 1741. 
\bibitem{benhar93}
O. Benhar and V.R. Pandharipande, Rev. Modern Phys. 65 (1993)
817. 
\bibitem{guardiola96} R. Guardiola, P.I. Moliner, J. Navarro,
R.F. Bishop, A. Puente and Niels R. Walet, Nucl. Phys. A 609
(1996) 218.
\bibitem{pandharipande79} V.R. Pandharipande and R.B. Wiringa,
Rev. Mod. Phys. 51 (1979) 821.
\bibitem{fantoni87}
S. Fantoni and V.R. Pandharipande, Nucl. Phys. A 473 (1987) 234. 
\bibitem{feldmeier98} H. Feldmeier, T. Neff, R. Roth and J. Schnack,
Nucl. Phys. A 632 (1998) 61. 
\bibitem{miller76} Gerald A. Miller and James E. Spencer, Annals of
Physics 100 (1976) 562.
\bibitem{marcvdh}
M. Vanderhaeghen, L. Machenil, J. Ryckebusch and M. Waroquier, 
Nucl. Phys. A 580 (1994) 551.
\bibitem{abbott} D. Abbott {\em et al.}, Phys. Rev. Lett. 80
(1998) 5072.
\bibitem{gearhart} C.C.Gearhart, PhD thesis, Washington University
(St. Louis, 1994), unpublished and W. Dickhoff, private communication.
\bibitem{blom98} 
K.I. Blomqvist {\em et al.}, Phys. Lett. B 421 
(1998) 71.
\bibitem{jan99} Jan Ryckebusch, Dimitri Debruyne, Wim Van Nespen and
Stijn Janssen, Phys. Rev. C 60 (1999) 034604. 
\bibitem{gercoprl2} 
C.J.G. Onderwater {\em et al.}, Phys. Rev. Lett.
81 (1998) 2213.
\bibitem{ronald99} R. Starink {em et al.}, in press and Ph.D. Thesis,
Free University of Amsterdam (1999), unpublished.
\bibitem{dimitrienlieven}
D. Van Neck, L. Van Daele, Y. Dewulf and M. Waroquier, Phys. Rev. C 56
(1997) 1398. 
\bibitem{leeuwe98}
J.J. van Leeuwe {\em et al.}, Nucl. Phys. A 631 (1998) 593c. 
\bibitem{templon99}
J.A. Templon, nucl-ex/9902008.
\bibitem{shalev} TJNAF experiment E97-106 ``Studying the Internal
Small-distance structure of Nuclei via the Triple Coincidence 
($e,e'p+N$) measurement'' (spokesperson E.~Piasetzky).
\bibitem{bianci95} A. Bianconi, S. Jeschonnek, N.N. Nikolaev and
B.G. Zakharov, Phys. Lett. B 363 (1995) 217.

\end{thebibliography}
\end{document}